\documentclass[sigplan,10pt,nonacm,screen]{acmart}

\usepackage{xspace}
\usepackage{subcaption}
\usepackage{algorithm}
\usepackage{algpseudocode}
\usepackage{pifont}
\usepackage{tabularx}
\usepackage{makecell}
\usepackage{multirow}
\usepackage{enumitem}

\makeatletter
\let\hyxmp@parse@acmart\relax
\makeatother

\AtBeginDocument{
  \hypersetup{
    colorlinks=true,
    linkcolor=blue,
    citecolor=magenta,
    filecolor=cyan,
    urlcolor=red,
    pdfauthor={Wenhao Chen and Hong Zhang},
    pdfpublisher={},
    pdfpubtype={},
    pdfpublication={},
    pdfvolumenum={},
    pdfissuenum={},
    pdfdoi={},
    pdfisbn={}
  }
}

\newcommand{\IState}{\State\hspace*{\algorithmicindent}}
\newcommand{\IIState}{\State\hspace*{\algorithmicindent}\hspace*{\algorithmicindent}}
\newcommand{\IIIState}{\State\hspace*{\algorithmicindent}\hspace*{\algorithmicindent}\hspace*{\algorithmicindent}}

\def\sys{{ConfluxServe}\xspace}
\def\para{{Conflux Parallelism}\xspace}
\def\inst{{XP instance}\xspace}
\def\overlap{{attention overlap}\xspace}

\newcommand{\parab}[1]{\paraspace\noindent{\bf #1} }

\setlist[itemize,1]{
    leftmargin=*,
    align=left,
    itemsep=2pt,
    topsep=2pt,
    parsep=0pt,
    partopsep=0pt
}
\setlist[enumerate,1]{
    leftmargin=*,
    align=left,
    itemsep=2pt,
    topsep=2pt,
    parsep=0pt,
    partopsep=0pt
}

\makeatletter
\patchcmd{\@xsect}{\clubpenalty\@M}{\clubpenalty\@clubpenalty}{}{}
\makeatother

\makeatletter
\renewcommand\section{\@startsection{section}{1}{\z@}%
  {-1.4ex \@plus -.1ex \@minus -.05ex}%
  {.25\baselineskip}%
  {\ACM@NRadjust\@secfont}}
\renewcommand\subsection{\@startsection{subsection}{2}{\z@}%
  {-1.1ex \@plus -.07ex \@minus -.04ex}%
  {.25\baselineskip}%
  {\ACM@NRadjust\@subsecfont}}
\makeatother
\makeatletter
\renewcommand\paragraph{\@startsection{paragraph}{4}{\z@}%
  {-0.4ex \@plus -.05ex \@minus -.03ex}%
  {-3.5\p@}%
  {\ACM@NRadjust{\bfseries\@adddotafter}}}
  \renewcommand{\parab}[1]{\paragraph{#1}}
\let\ACM@origsection\section
\let\ACM@origsubsection\subsection
\let\ACM@origparagraph\paragraph
\makeatother

\begin{document}

\title[ConfluxServe]{{\fontsize{18pt}{20pt}\selectfont
\sys: Reconciling Resource Utilization
and Parallelism Efficiency across
Prefill and Decode in LLM Serving
}}


\author{Wenhao Chen}
\affiliation{%
  \institution{University of Waterloo}
  \city{Waterloo}
  \country{Canada}
}
\email{wenhao.chen@uwaterloo.ca}

\author{Hong Zhang}
\affiliation{%
  \institution{University of Waterloo}
  \city{Waterloo}
  \country{Canada}
}
\email{hong.zhang@uwaterloo.ca}

\renewcommand{\shortauthors}{Chen and Zhang}

\begin{abstract}
Prefill and decode are complementary in resource usage, but often divergent in parallelism preferences. Executing the two phases separately preserves their preferred parallelism but leaves complementary resources underutilized, while co-batching improves utilization but often requires a uniform parallel configuration that hurts execution efficiency.

To resolve this utilization--efficiency dilemma, we propose \para\ (XP), a parallel execution abstraction that co-executes prefill and decode with \textit{different} parallelism within the \textit{same} batch. \para exploits operator-level Transformer structure to construct a \textit{heterogeneous} execution path with different batching and parallelism strategies across operators, combining the best of co-batching and phase-specific execution. Building on \para, we design \sys, an SLO-aware serving system that maximizes cluster goodput under TTFT and TBT constraints. Evaluation on production workloads shows that \sys improves serving goodput %
by $1.26$--$1.62\times$ over the strongest baselines.
\end{abstract}

\hypersetup{urlcolor=black}
\maketitle
\hypersetup{urlcolor=red}
\pagestyle{plain}

\vspace{-0.05in}
\section{Introduction}

Large language models (LLMs) are becoming core infrastructure for modern applications, from code generation~\cite{openai2026codex,anthropic2026claudecode} and customer-support assistants \cite{wulf2024customerservice} to scientific copilots ~\cite{gottweis2026coscientist}. Serving these applications requires handling large volumes of concurrent inference requests, each subject to stringent latency objectives such as time-to-first-token (TTFT) and time-between-tokens (TBT), both of which are critical to user experience.
{Accordingly, LLM serving systems seek to maximize \textit{goodput}: the maximum sustainable request rate under TTFT and TBT SLO constraints~\cite{zhong2024distserve,wu2024loongserve,qin2025mooncake}.}

LLM inference consists of two distinct phases: prefill and decode. Prefill processes the input prompt to produce the first output token and populate the KV cache, whereas decode generates subsequent tokens iteratively by reusing the cached state. Maximizing goodput across these two phases is challenging because they differ not only in their resource bottlenecks, but also in their preferred parallelism configurations under SLO constraints.

From a resource-utilization perspective, prefill and decode are naturally \textit{complementary}. Prefill is predominantly compute-intensive, whereas decode is constrained primarily by memory bandwidth and KV-cache capacity. This complementarity motivates \textit{co-execution} designs such as co-batching (\textit{e.g.,} {Chunked Prefill}~\cite{agrawal2024taming}) and spatial multiplexing~\cite{chen2026towards,lin2026bullet}, which execute the two phases concurrently in the same serving instance to improve overall hardware utilization (\S\ref{subsec:p1}).

From a parallelism perspective, however, prefill and decode impose fundamentally \textit{divergent} parallelism requirements under their respective SLOs. Long prefills must scale out to satisfy TTFT and favor an efficient scale-out configuration, whereas decode can typically satisfy TBT on a much narrower execution group, with further scale-out introducing additional communication and reducing execution efficiency.
This divergence favors \textit{phase-isolated} execution,
which separates prefill and decode either across  machines~\cite{zhong2024distserve,patel2024splitwise,li2026tetris} or over time~\cite{wu2024loongserve,du2025ecoserveenablingcosteffectivellm},
allowing each to use a parallelism configuration better suited to its execution characteristics (\S\ref{subsec:p2}).

The two observations together create a dilemma: co-executing prefill and decode improves hardware utilization by exploiting their complementary resource demands, but typically requires them to share a common parallelism configuration, forcing at least one phase away from its efficient operating point. In contrast, phase-isolated execution allows each phase to use a more suitable parallelism configuration, but forgoes the utilization benefits of co-execution. We refer to this tension as the \textit{utilization--efficiency dilemma} (\S\ref{subsec:existing-designs}).

Is it possible to exploit prefill--decode resource complementarity without sacrificing phase-specific parallelism efficiency? We answer affirmatively with {\para} (XP), a parallel execution abstraction that co-executes prefill and decode with \textit{different} parallelism within the \textit{same} batch.

\para is enabled by the observation that both the feasibility and benefit of co-batching computations with different parallelism configurations are \textit{operator-dependent}.  Dense linear operators benefit strongly from co-batching and remain co-batchable even when prefill and decode use different degrees of sequence parallelism (SP). Attention, in contrast, cannot be directly co-batched within a single invocation under different parallelism configurations, yet benefits much less from  co-batching in the first place.

This observation motivates the central design choice in \para: cross-phase parallel execution is composed at the  \textit{operator level} rather than uniformly across the model. %
Concretely, \para fully co-batches the linear computations of prefill and decode, gracefully accommodating their distinct SP degrees.
At attention, \para executes the two phases under distinct parallelism configurations, with decode attention remaining local and prefill attention executing across ranks with SP.
This heterogeneous execution combines three benefits that are difficult to achieve simultaneously. First, \para preserves efficient co-batching through the dense linear operators, exploiting resource complementarity. Second, it allows each phase to retain its preferred parallelism configuration \textit{throughout execution}, avoiding the efficiency loss of uniform parallelism.
Third, it enables coordinated spatial multiplexing between prefill and decode attention within a {narrow}, repeatable window, further exploiting complementary resources with stable cross-phase interference (\S\ref{sec:novel-parallelism}).

Building on \para, we design \sys, an SLO-aware serving system that coordinates heterogeneous prefill--decode execution to maximize goodput under both TTFT and TBT constraints (\S\ref{sec:overview}). \sys keeps ongoing decodes localized to their assigned ranks while co-executing high-SP prefill chunks across ranks using the residual capacity left by decode.
In each execution iteration, %
\sys prioritizes pending prefills by TTFT urgency and dispatches their tokens across ranks to maximize admitted prefill work without violating decode TBT SLOs.
Once a prefill completes, \sys selects its decode-residency rank to maximize co-execution opportunities throughout its decode lifetime.

\begin{figure}[tbp]
    \centering
    \includegraphics[width=0.7\linewidth]{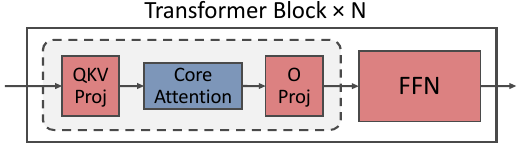}
    \caption{Transformer block workflow in LLM inference.}
\label{fig:llm-steps}
\end{figure}

\sys comprises four key components.
First, to execute heterogeneous batches efficiently, \sys retains CUDA Graph {optimization} for shared linear operators and limits interference during attention multiplexing through
SM partitioning (\S\ref{sec:engine-execution}).
Second, to ensure heterogeneous co-execution respects TBT SLOs, we develop a latency model capturing critical-rank and critical-path dependencies, and a profile-fitted estimator for co-batched iteration latency (\S\ref{sec:estimator}). Third, to maximize admissible prefill work across current and future iterations, \sys uses waterfill prefill dispatch to reduce current rank imbalance, and dual-mode decode placement to balance future execution efficiency against KV capacity (\S\ref{sec:dispatcher}).
Finally, \sys integrates TTFT-aware admission and prioritization with
dispatch and placement
in
{an}
end-to-end scheduler to maximize goodput (\S\ref{sec:scheduler}).

We implement \sys on vLLM (\S\ref{sec:implementation}) and evaluate it on representative workloads against state-of-the-art baselines (\S\ref{sec:evaluation}). Under FCFS scheduling, \para sustains $1.33$--$1.62\times$ higher goodput than the strongest existing designs. With its full scheduler, \sys improves goodput by $1.26$--$1.52\times$ over the strongest baselines, even after augmenting each with enhanced scheduling. These gains remain consistent across model configurations and workloads.

\section{Background}
\label{sec:bg-llm}

\parab{Computation Workflow in LLM Inference.}
Most LLMs~\cite{grattafiori2024llama3,yang2025qwen3} consist of stacked Transformer blocks~\cite{vaswani2017attention}, as illustrated in Figure~\ref{fig:llm-steps}.
Each block applies a QKV projection, performs core attention (\emph{attention} hereafter), and then applies an output projection followed by a feed-forward network (FFN).
QKV projection, output projection, and FFN consist primarily of dense linear operators implemented using general matrix-matrix multiplication (GEMM) kernels.
In contrast, attention combines matrix multiplications, softmax, and element-wise operations, and is typically implemented using specialized fused kernels such as FlashAttention~\cite{dao2022flashattention}.
Attention accesses keys and values of preceding tokens, which inference systems retain in a KV cache to avoid recomputation~\cite{pope2023efficiently,kwon2023efficient}.

\parab{Prefill and Decode.}
LLM inference consists of two phases. During prefill, the system processes the input tokens in parallel, performs causal attention over the prompt, and materializes their KV cache. Prefill phase is typically compute-intensive,
and its latency is measured as time-to-first-token (TTFT).
During decode, the system generates tokens autoregressively, processing one token per iteration while reading the historical KV cache. Decode phase is predominantly memory-bound~\cite{pope2023efficiently,patel2024splitwise}, and its inter-token latency is measured by time-between-tokens (TBT).
 LLM serving typically imposes SLOs on both TTFT and TBT. TTFT targets are commonly on the order of a few seconds, while TBT targets typically range from tens to a few hundred milliseconds~\cite{zhong2024distserve,zhang2026jitserve}.

\begin{figure}[tb]
    \centering
    \includegraphics[width=1.0\linewidth]{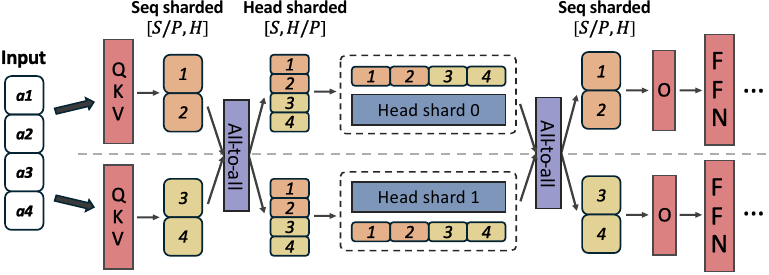}
    \caption{
    {Ulysses sequence parallelism.}
    }
\label{fig:ulysses}
\end{figure}

\parab{Parallelism Strategies.}
LLM serving systems employ different parallelism strategies to scale inference across GPUs.
\emph{Tensor parallelism} (TP)~\cite{shoeybi2020megatronlm} shards model weights across ranks, reducing per-rank computation and memory usage.
\emph{Pipeline parallelism} (PP)~\cite{huang2019gpipe} partitions consecutive blocks across stages, reducing per-GPU model memory but not directly reducing single-request latency, and may suffer from pipeline bubbles.
\emph{Sequence parallelism} (SP)~\cite{fang2024usp} partitions the sequence dimension across ranks, reducing per-rank computation while replicating model weights.

{Two common SP designs are} Ring-SP~\cite{liu2023ring} and Ulysses-SP~\cite{jacobs2023deepspeedulyssesoptimizationsenabling}.
Ring-SP keeps the sequence partitioned and circulates KV blocks across ranks, whereas Ulysses-SP uses \texttt{All-to-All} to transpose between sequence- and head- partitioned layouts around attention.
We focus on Ulysses-SP because it achieves more favorable communication scaling and better supports chunked-prefill execution.
{Appendix~\ref{app:parallel} details the execution characteristics of Ring-SP, Ulysses-SP, and  TP.}
Our analysis and heterogeneous-parallelism design are not specific to Ulysses-SP and also apply to Ring-SP.
Unless otherwise stated, SP refers to Ulysses-SP in the paper.

Figure~\ref{fig:ulysses} illustrates Ulysses-SP for sequence length $S$, SP degree $P$, and hidden dimension $H$.
{Each rank initially holds $S/P$ tokens with the full hidden dimension $H$.}
{The first \texttt{All-to-All} transposes the QKV states from sequence partitioning to head partitioning, so that each rank holds the full sequence for a subset of attention heads.}
After computing attention on these head shards, a second \texttt{All-to-All} restores the sequence-partitioned layout for subsequent operators.

\section{The Utilization–Efficiency Dilemma}
\label{sec:dilemma}
We observe a key tension in LLM serving: prefill and decode exhibit complementary resource demands (\S\ref{subsec:p1}) but divergent parallelism preferences (\S\ref{subsec:p2}). Together, they create a utilization--efficiency dilemma for existing designs (\S\ref{subsec:existing-designs}).

\subsection{Prefill--Decode Resource Complementarity}
\label{subsec:p1}

\begin{figure}[tb]
    \centering
    \includegraphics[width=1\columnwidth]{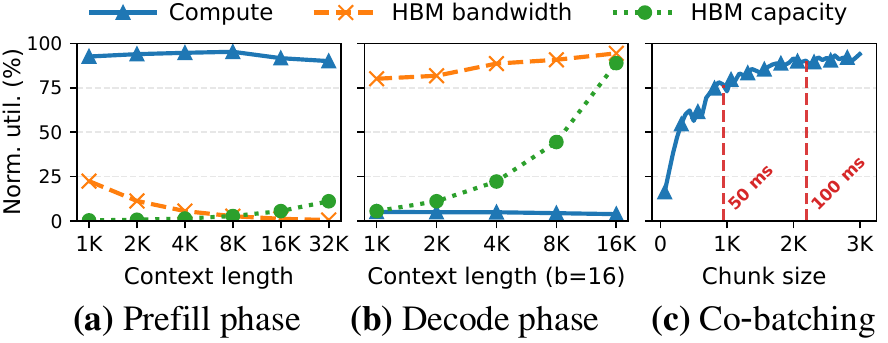}
    \caption{
    \textbf{Prefill--decode resource complementarity} (Qwen3-14B/H100):
    (a--b) Normalized compute, HBM-bandwidth, and HBM-capacity utilization for prefill and decode.
    (c) Normalized compute utilization when co-batching prefill with 16 decodes at 8K context; dashed lines mark the largest chunks satisfying 50\,ms and 100\,ms TBT targets.
    }
    \label{fig:phase-utilization}
\end{figure}

Prefill and decode stress different GPU resources. Prefill processes prompt tokens in parallel and is primarily compute-intensive, whereas decode generates one token per request per iteration while repeatedly accessing model weights and resident KV caches, making it primarily constrained by HBM bandwidth and KV-cache capacity. %
We validate this contrast by profiling {Qwen3-14B on an NVIDIA H100} over varying prefill lengths and decode context lengths. Figure~\ref{fig:phase-utilization}(a--b) shows that prefill sustains high compute utilization, while decode approaches the HBM-bandwidth limit and consumes increasing KV capacity as context grows.

This complementarity creates an opportunity for co-execution.
{Chunked Prefill~\cite{agrawal2024taming} exploits it by splitting long prefills into chunks and co-batching them with ongoing decodes, using the remaining TBT headroom for compute-intensive prefill work.}
In our profiling, a batch of 16 decodes with 8K-token contexts takes only about 20\,ms per iteration. As Figure~\ref{fig:phase-utilization}(c) shows, approximately 1K and 2.2K prefill tokens can therefore be co-batched while satisfying 50\,ms and 100\,ms TBT targets, increasing normalized compute utilization from about 16\% to 75\% and 90\%, respectively. Another line of work~\cite{chen2026towards,lin2026bullet} exploits this complementarity through \textit{spatial multiplexing},
{partitioning GPU SMs between concurrently executing prefill and decode kernels.}

\subsection{Prefill–Decode Parallelism Divergence}
\label{subsec:p2}
We observe that prefill and decode favor fundamentally different parallelism configurations with their respective SLOs.

\parab{Prefill requires scale-out.}
A prefill request must process its entire prompt before producing the first token, making the TTFT SLO increasingly difficult to satisfy as prompt length grows.
Moreover, production LLM workloads often exhibit heavy-tailed prompt length distributions. For example, P95 prompt lengths reach 26K and 49K in Mooncake-Toolagent and Mooncake-Synthetic {(Table~\ref{tab:workloads})}.

\begin{figure}[tb]
    \centering
    \includegraphics[width=1\linewidth]{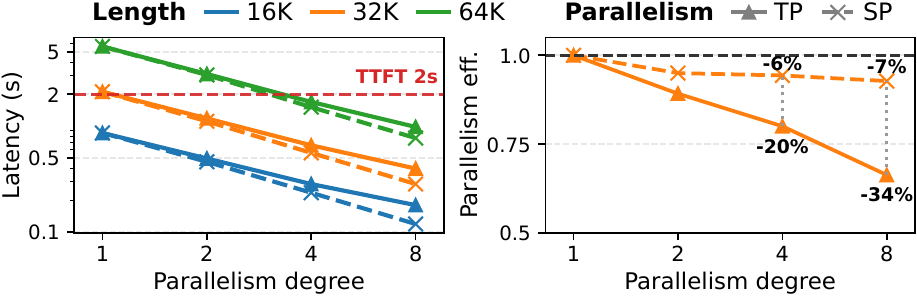}
    \caption{
    \textbf{Prefill scaling} (Qwen3-14B/H100).
    (a) Execution latency for 16K--64K prompts.
    (b) Parallelism efficiency under TP and SP for a {32K} prefill.
    }
\label{fig:prefill-parallelism}
\end{figure}

Figure~\ref{fig:prefill-parallelism}(a) profiles the latency of 16K, 32K, and 64K prefills on Qwen3-14B across different parallelism degrees.
{To sustain a 2\,s TTFT SLO~\cite{zhang2026jitserve}},
32K prompts require at least two GPUs, while 64K prompts require at least four. These degrees are only \textit{lower bounds}
{derived from isolated execution latency.}
In practice, additional scale-out is needed to accommodate queueing delay under traffic variability.

Figure~\ref{fig:prefill-parallelism}(b) further examines
{how efficiently TP and SP realize this scale-out.}
Given a workload, we define the \textit{parallelism efficiency} of a parallelism strategy at degree $d$ as
\begin{equation}
E(d) = {T(1)}/{(d \cdot T(d))},
\end{equation}
\noindent
where $T(d)$ denotes execution latency on $d$ GPUs, and $E(d)=1$ corresponds to ideal linear scaling. While both TP and SP reduce prefill latency, SP maintains substantially higher parallelism efficiency, retaining {93\%} at degree 8 versus {66\%} for TP. This higher efficiency stems from SP's better communication scaling (analyzed in Appendix~\ref{app:parallel}).
We observe a similar %
trend across prompt lengths.

\noindent{\underline{\textit{\textbf{Takeaway 1:}}}} \textit{Long-context prefill requires scale-out to satisfy TTFT, and {often} favors SP for higher parallelism efficiency.}

\parab{Decode favors narrow execution.}
Decode exhibits a different scaling behavior. Each iteration processes only one new token per request rather than the entire {prompt}
as in prefill.
Moreover, decode operates on batches, and its latency %
reflects the context-length mix of the batch
rather than the long-context tail of an individual request.
For example, median prompt lengths are 6K / 12K tokens in Mooncake-Toolagent / Mooncake-Synthetic, respectively, substantially below their P95 lengths.
In addition, {KV-cache capacity further bounds the resident decode requests and, consequently, per-iteration decode latency.}
Together, decode often requires only a narrow parallelism configuration to satisfy TBT SLO.

\begin{figure}[tb]
    \centering
    \includegraphics[width=1\linewidth]{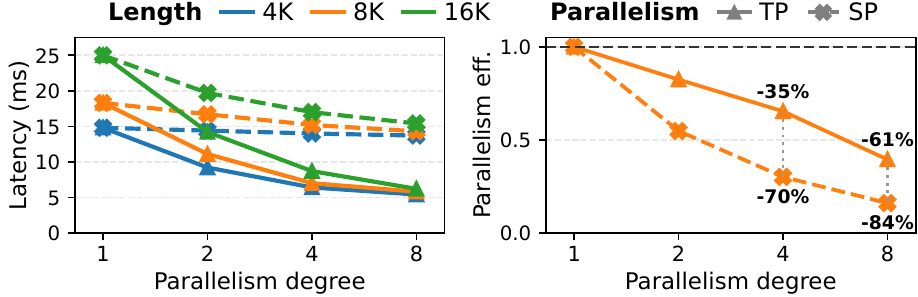}
    \caption{
    \textbf{Decode scaling}  (Qwen3-14B/H100).
  (a) Decode execution latency (batch size 16, 4K--16K contexts).
    (b) Parallelism efficiency under TP and SP (batch size 16, 8K context).
    }
\label{fig:decode-parallelism}
\end{figure}

Figure~\ref{fig:decode-parallelism}(a) profiles decode batches of 16 requests each with 4K, 8K, and 16K contexts on Qwen3-14B across parallelism degrees.
Further increasing batch size or context length exceeds single-GPU HBM capacity. %
A single-GPU configuration already completes each iteration within 25\,ms across these settings, leaving substantial margin even {under a tight 50\,ms TBT SLO~\cite{goel2025niyama}.}
Figure~\ref{fig:decode-parallelism}(b) further shows that scaling out decode is inefficient: parallelism efficiency drops to only 39\% and 16\% at degree 8 under TP and SP, respectively.
This low efficiency stems from decode's limited per-iteration work, which provides little opportunity to amortize the communication overhead of scale-out.
We observe a similar trend across context lengths.

\noindent{\underline{\textit{\textbf{Takeaway 2:}}}}
\textit{Decode can typically satisfy TBT on a narrow parallelism configuration, with further scale-out substantially reducing parallelism efficiency. }

\parab{Uniform-parallelism tax.}
The two takeaways together create an inherent tradeoff when prefill and decode are forced to share the same parallelism configuration.
A low degree preserves decode parallelism
efficiency but can leave long prefills unable to meet TTFT, whereas a high degree
{improves prefill TTFT feasibility}
at the cost of unnecessary communication and lower decode  parallelism
efficiency.
Thus, any uniform configuration must compromise between prefill SLO feasibility and decode  efficiency. We refer to this loss from enforcing a common parallelism configuration as \textit{uniform-parallelism tax}.
Such a tax makes phase-isolated designs attractive: separating prefill and decode either across
machines~\cite{zhong2024distserve,patel2024splitwise,li2026tetris} or over time~\cite{wu2024loongserve,du2025ecoserveenablingcosteffectivellm} allows each phase to use its preferred parallelism configuration. %

\subsection{Limitation of existing works}
\label{subsec:existing-designs}

The preceding observations expose two conflicting requirements: resource complementarity favors prefill--decode co-execution for hardware utilization (\S\ref{subsec:p1}), whereas parallelism divergence favors phase-isolated execution for parallelism efficiency (\S\ref{subsec:p2}).
As summarized in Table~\ref{tab:pd-taxonomy},
existing designs satisfy one requirement at the expense of the other, creating what we {term} the \emph{utilization--efficiency dilemma}.

\parab{Phase-Isolated Designs.}
Static isolation designs~\cite{zhong2024distserve,patel2024splitwise,li2026tetris} dedicate separate GPU pools to the two phases, enabling independent provisioning but potentially stranding resources when their demands are imbalanced. Elastic isolation designs~\cite{wu2024loongserve,du2025ecoserveenablingcosteffectivellm} dynamically reassign GPU instances between prefill and decode over time, thereby reducing resource imbalance compared with static isolation.
Both approaches keep prefill and decode isolated in \textit{separate} iterations. This enables phase-specific parallelism, but prevents the two phases from exploiting their complementary {compute and HBM-bandwidth demands.}

\begin{table}[tb]
\centering
\small

\begin{tabularx}{\columnwidth}
{@{}X>{\centering\arraybackslash}m{0.22\columnwidth}>{\centering\arraybackslash}m{0.22\columnwidth}@{}}
\toprule
\noalign{\vskip -10pt}

\\[-11pt]
\raisebox{-1pt}{\fontsize{9pt}{10pt}\selectfont\parbox[c][2em][c]{\linewidth}{\hspace*{0em}\textbf{Design}}} &
{\fontsize{8pt}{10pt}\selectfont\hspace*{0em}\makecell{\textbf{Resource}\\ \textbf{Complementarity}}} &
\raisebox{2pt}{\fontsize{8pt}{10pt}\selectfont\makecell{\textbf{Phase-Specific}\\ \textbf{Parallelism}}}
\\[-7pt]
\midrule

\multicolumn{3}{@{}l}{\textsc{Phase-isolated designs}} \\
 \specialrule{\lightrulewidth}{0.5pt}{1.5pt}
\makecell[l]{
    $\circ$ Static isolation \cite{zhong2024distserve,patel2024splitwise,li2026tetris} \\
    \quad {\fontsize{8.5pt}{10pt}\selectfont \textit{Representative:} \textit{DistServe}~\cite{zhong2024distserve}}
}
& \hspace*{1em}\ding{55} & \ding{51} \\
\makecell[l]{
    $\circ$ Elastic isolation \cite{wu2024loongserve,du2025ecoserveenablingcosteffectivellm} \\
    \quad {\fontsize{8.5pt}{10pt}\selectfont \textit{Representative:} \textit{LoongServe}~\cite{wu2024loongserve}}
}
& \hspace*{1em}\ding{55} & \ding{51} \\
[-6pt]
 \specialrule{\lightrulewidth}{0.5pt}{1.5pt}

\multicolumn{3}{@{}l}{\textsc{Co-execution designs}} \\
 \specialrule{\lightrulewidth}{0.5pt}{1.5pt}
\makecell[l]{
    $\circ$ Co-batching ~\cite{agrawal2024taming,wang2026taichi,holmes2024deepspeedfastgen,ruan2026libra}\\
    \quad {\fontsize{8.5pt}{10pt}\selectfont \textit{Representative}: \textit{Chunked Prefill}~\cite{agrawal2024taming}}
}
& \hspace*{1em}\ding{51} & \ding{55} \\
[-5pt]
\makecell[l]{
    $\circ$ Spatial multiplexing~\cite{chen2026towards,lin2026bullet,feng2025windserve} \\
    \quad {\fontsize{8.5pt}{10pt}\selectfont \textit{Representative}: \textit{MuxWise}~\cite{chen2026towards}}
}
& \hspace*{1em}\ding{51} & \ding{55} \\
[-5pt]
\specialrule{\lightrulewidth}{0pt}{3pt}

\hspace*{0.5em}\textbf{\para (Ours)}
& \textbf{\hspace*{1em}\ding{51}}
& \textbf{\ding{51}} \\
\bottomrule
\end{tabularx}

\caption{Utilization--efficiency dilemma in existing designs.}
\label{tab:pd-taxonomy}

\end{table}

\parab{Co-Execution Designs.}
Co-batching {designs}~\cite{agrawal2024taming} mix prefill chunks with ongoing decodes in the same batch; recent designs~\cite{ruan2026libra,wang2026taichi} further {adapt} how prefill and decode work are partitioned and co-batched according to workload conditions and SLOs. Spatial-multiplexing {designs}~\cite{chen2026towards,lin2026bullet} instead partition a GPU's SMs between prefill and decode so that their kernels execute concurrently. Both approaches exploit prefill--decode resource complementarity on the same GPUs. However, they typically require the two phases to share a common parallelism configuration, incurring the \emph{uniform-parallelism tax} identified in \S\ref{subsec:p2}.

\parab{Example.}
We illustrate the dilemma on four H100 GPUs on Qwen3-14B using {Mooncake-Toolagent workload.}
Following \S\ref{subsec:p2}, we set the prefill scale-out degree to 4 to maintain sufficient TTFT headroom, while degree 1 leaves
sufficient
TBT slack for decode. We maintain {96} ongoing decode requests while progressively increasing the offered prefill load, and report the maximum prefill throughput each design sustains without violating the decode TBT SLO.

\begin{figure}[tb]
    \centering
    \includegraphics[width=0.96\linewidth]{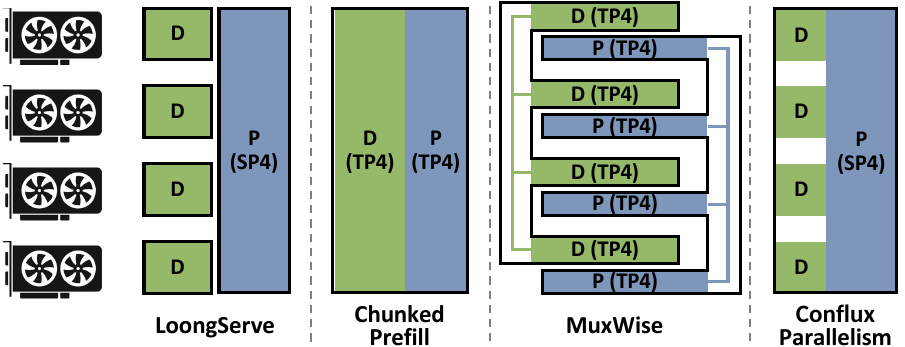}
    \caption{
Representative systems under their designed parallelism configurations.  Each black box denotes a single batch.  %
    }
\label{fig:toy-example}
\end{figure}

Figure~\ref{fig:toy-example} instantiates representative systems using their designed parallelism configurations.\footnote{We omit static phase isolation because prefill requires all four GPUs for scale-out, leaving no separate decode pool.}
LoongServe~\cite{wu2024loongserve} (dynamic isolation) preserves phase-specific parallelism by running prefill at SP4 and decode at SP1, but temporally isolates the two phases, leaving substantial compute unused during decode execution.
Chunked Prefill~\cite{agrawal2024taming} (co-batching) exploits resource complementarity by co-batching prefill and decode under a uniform TP4 configuration. However, the high decode parallelism lowers parallelism efficiency and consumes more TBT headroom, limiting the admissible prefill chunk size. %
MuxWise~\cite{chen2026towards} (spatial multiplexing) partitions GPU SMs between
concurrent prefill and decode kernels using a shared TP4 configuration. It therefore still incurs the uniform-parallelism tax,
while SM partitioning further limits the compute capacity available to prefill to protect decode latency.
Moreover, because the two phases execute as separate kernels, their linear operators cannot be co-batched to amortize model-weight traffic, and the concurrent kernels can contend for shared memory and communication resources.

Table~\ref{tab:perf-comparison} reports the prefill throughput of each design under 50\,ms and 100\,ms TBT SLOs, alongside \para~(ours), which unlocks {up to $1.40\times$ higher prefill throughput} by addressing the dilemma. We show how \para realizes this opportunity next.

\begin{table}[tb]
\centering
\resizebox{\columnwidth}{!}{%
\begin{tabular}{@{}lcccc@{}}
\toprule
\noalign{\vskip -2pt}
\small\textbf{TBT SLO} & \small\textbf{LoongServe} & \makecell{\small\textbf{Chunked}\\\small\textbf{Prefill}}
& \small\textbf{MuxWise} & \makecell{\small\textbf{Conflux}\\\small\textbf{Parallelism}} \\[-2pt]
\midrule
\noalign{\vskip -1pt}
50\,ms  & 2.41 & 4.17 & 4.38 & \textbf{6.15} \\
100\,ms & 5.17 & 5.92 & 5.43 & \textbf{8.27} \\[-1pt]
\bottomrule
\end{tabular}%
}
\caption{Prefill throughput (req./s) under different TBT SLOs. }
\label{tab:perf-comparison}
\end{table}

\section{\para}
\label{sec:novel-parallelism}

\para is enabled by an operator-level asymmetry in \textit{cross-parallelism co-batching}: Linear operators and attention differ in the feasibility and benefit of co-batching computations with different parallelism configurations.

{\textit{Dense linear operators}}, including QKV and output projections and FFNs, apply the same weights independently to each token. Co-batching prefill and decode therefore amortizes model-weight traffic over more tokens, improving weight reuse and arithmetic intensity. Crucially, such co-batching only requires a shared weight matrix and feature dimension.
This creates an opportunity to co-batch prefill and decode under \textit{different SP} degrees: SP only changes how the sequence dimension is partitioned across ranks, while leaving the feature dimension and model weights unchanged. Each rank can therefore concatenate its local prefill and decode tokens and process them in a single GEMM.

\textit{Attention} exhibits the opposite  profile.
{Attention computations can be directly co-batched only when they share a common QKV layout.}
Different parallelism configurations with different %
QKV sharding therefore cannot be co-batched within a single invocation.
Yet attention gains much less from cross-parallelism co-batching in the first place:
each request attends to its own KV state, so batching more requests increases computation and KV-state traffic proportionally without asymptotically increasing arithmetic intensity. Appendix~\ref{app:op-analysis} details the operator analysis.

\begin{figure}[tb]
    \centering
    \includegraphics[width=1.0\columnwidth]{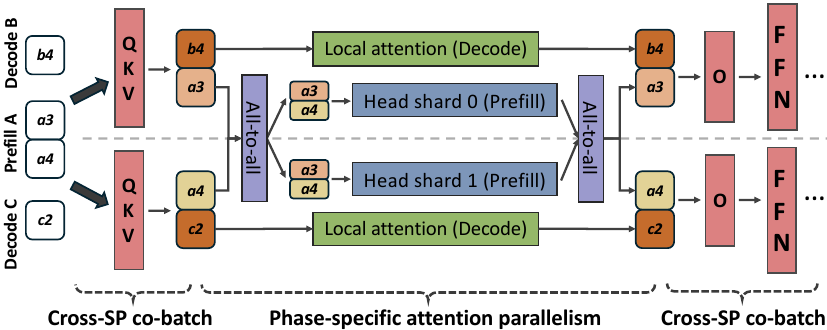}
    \caption{
    Operator-level execution flow of \para across two ranks. %
    Prefill and decode with different SP degrees are co-batched through the shared linear operators while retaining distinct attention-parallelism configurations. %
    }
    \label{fig:novel-parallelism}
\end{figure}

\parab{\para execution flow.}
Figure~\ref{fig:novel-parallelism} illustrates how \para exploits this asymmetry to enable co-batching while preserving each phase’s parallelism configuration throughout the Transformer layer.
Prefill request $A$ processes a chunk containing $a_3$ and $a_4$, while decode requests $B$ and $C$ process tokens $b_4$ and $c_2$, respectively. \para partitions $A$'s prefill chunk across the two ranks with SP degree 2, while executing $B$ and $C$ locally, one on each rank.
Before attention, each rank co-batches its local prefill and decode tokens, forming $[a_3,b_4]$ on rank 0 and $[a_4,c_2]$ on rank 1. %
Each rank then processes its concatenated tokens together in a single QKV-projection.
At attention, the tokens follow their retained layouts concurrently:
decode attention remains local, while prefill attention executes with SP using two \texttt{All-to-All} collectives and restores sequence partitioning afterward.
After attention, both phases again expose compatible rank-local token tensors, allowing each rank to co-batch them through the remaining linear operators.

\begin{figure}[tb]
    \centering
    \includegraphics[width=0.96\columnwidth]{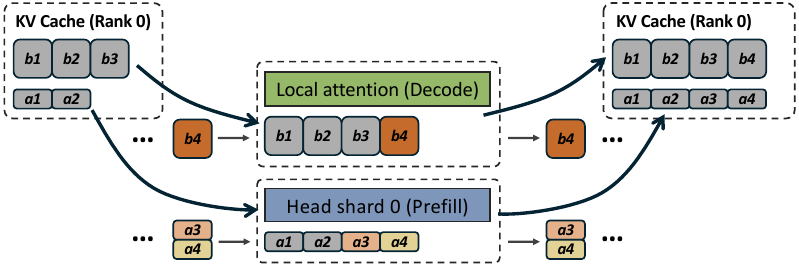}
    \caption{
    KV residency in \para, illustrated on \textit{one rank}. Decode KV remains local, prefill KV remains head-sharded, and cross-layout KV transfer is required only at the prefill-to-decode transition.
    }
    \label{fig:novel-kv-cache}
\end{figure}

\parab{KV residency and KV transfer.}
As illustrated in Figure~\ref{fig:novel-kv-cache}, \para preserves the KV-residency layout of each phase. Decode request $B$ extends its rank-local KV cache with $b_4$ without inter-rank communication, while prefill request $A$ appends $a_3$ and $a_4$ to its head-sharded KV cache under SP.
Diverging and reconverging the attention paths therefore introduces no additional KV redistribution: decode KV remains local, while prefill KV remains head-sharded.
This head-sharded prefill layout also naturally supports chunked prefill, allowing subsequent chunks to reuse historical KV without redistribution.
The only additional KV transfer occurs when a request transitions from distributed prefill to localized decode, at which point its KV shards are transferred once to the selected decode rank. This transfer uses the intra-node high-bandwidth interconnect and can be effectively overlapped with execution of the final prefill chunk (\S\ref{sec:implementation}).

\parab{\para over a base TP configuration.}
The preceding discussion assumes one GPU per logical \para rank. For larger models, \para composes naturally with TP by treating each fixed TP group as one logical rank: prefill applies SP across logical ranks, while each decode request remains localized within one rank. Since both phases share the same base-TP layout, their rank-local tokens can still be co-batched through TP-sharded linear operators and their associated TP collectives, while attention retains phase-specific parallelism across logical ranks. Appendix~\ref{app:tp-sp} details the base-TP execution workflow. %

\parab{Benefits.} \para provides three key benefits:

\vspace{0.01in}
\noindent{\textit{\textbf{1. Resource complementarity via co-batching.}}} \para co-batches prefill and decode in shared linear operators, combining their complementary resource demands and amortizing model-weight traffic across the {mixed} batch.

\vspace{0.01in}
\noindent{\textit{\textbf{2. Phase-specific parallelism via heterogeneous attention.}}}  \para lets prefill use distributed attention while decode remains localized, preserving each phase's preferred parallelism and avoiding unnecessary decode communication and the associated parallelism-efficiency loss.

\vspace{0.01in}
\noindent {\textit{\textbf{3. {Coordinated}
multiplexing via attention overlap.}}}
\para enables coordinated spatial multiplexing of heterogeneous prefill and decode attention, further exploiting resource complementarity while keeping interference stable.
The overlap is \textit{stable}: the two paths diverge at a fixed point and reconverge immediately afterward, creating a narrow, repeatable overlap window.
It is also \textit{complementary}: communication- and compute-intensive prefill path overlaps with memory-bandwidth-intensive decode attention~\cite{zhu2025nanoflow}.
In contrast, request-level spatial multiplexing~\cite{chen2026towards,lin2026bullet} lets separate prefill and decode kernels progress independently throughout execution without an operator-bounded overlap window, leading to higher and less stable interference.

\section{\sys: Overview}
\label{sec:overview}

Building on \para, we design \sys, an SLO-aware serving system that coordinates heterogeneous prefill--decode execution to maximize serving goodput under TTFT and TBT constraints. \sys keeps ongoing decode requests localized to their assigned ranks and dispatches high-SP prefill chunks across ranks according to TTFT urgency. These prefill chunks co-execute with decode, utilizing its residual capacity without violating TBT SLOs.

\parab{XP Instances.}
We partition the GPU cluster into independent XP instances, which serve as the basic units of scheduling and execution. Within each instance, GPUs are organized into fixed base-TP groups, with each TP group forming one logical rank.
The base TP degree is chosen to accommodate model weights while reserving sufficient KV-cache capacity, and the remaining intra-node parallelism is used to maximize the SP degree.
Prefill executes across these logical ranks using the fixed SP configuration, while each decode request remains localized within one rank.
Although higher SP introduces additional communication, \para overlaps the distributed prefill path with local decode attention, mitigating the resulting execution overhead and {enabling near-linear} prefill scaling {when co-executed with decode} (\S\ref{sec:eval-parallelism-degree}).
Moreover, the fixed configuration avoids runtime resizing and KV redistribution. It also simplifies online scheduling by removing SP degree from the decision space.

 \begin{figure}[tb]
    \centering
    \includegraphics[width=0.96\columnwidth]{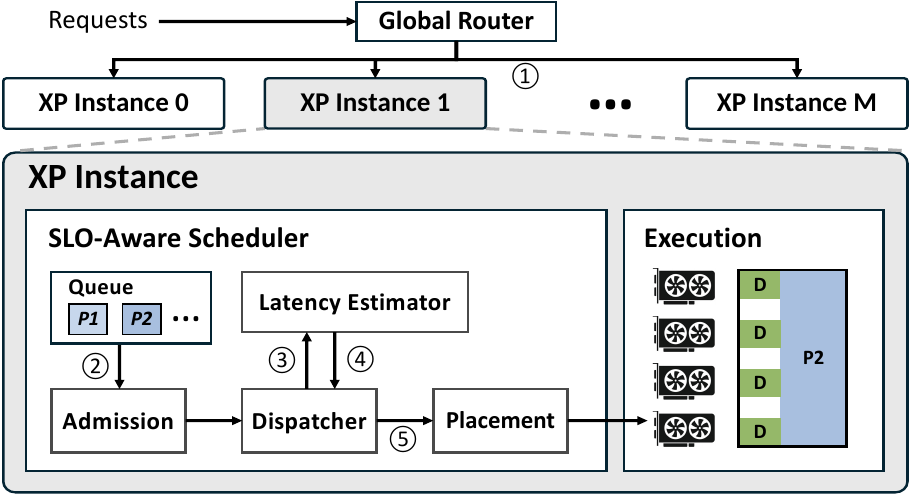}
    \caption{System overview of \sys. }
    \label{fig:system-overview}
\end{figure}

\parab
{Scheduling workflow.}
As illustrated in Figure~\ref{fig:system-overview}, newly arrived requests are routed to the XP instance with the lowest normalized queued workload, measured as queued work relative to the instance's processing capacity (\raisebox{-0.15ex}{\ding{172}}).
{
Within each instance, \sys organizes execution into serving iterations. Each iteration advances all resident decode requests by one token on their assigned ranks, ensuring stall-free decode. The remaining capacity available for prefill without violating the TBT SLO defines the \emph{TBT-safe prefill budget}.
}
The scheduler performs TTFT-aware admission and prioritization to determine which pending prefills consume this capacity and in what order (\raisebox{-0.15ex}{\ding{173}}).
Following this order, the waterfill dispatcher queries the latency estimator for candidate prefill allocations (\raisebox{-0.15ex}{\ding{174}}). Using these estimates, the dispatcher
{
determines the chunk size for each request and its corresponding per-rank assignment to maximize the TBT-safe prefill budget (\raisebox{-0.15ex}{\ding{175}}).
}
If a selected chunk completes the request's prefill, the scheduler invokes decode placement to select and reserve its decode rank (\raisebox{-0.15ex}{\ding{176}}).
Preselecting this rank allows the KV transfer to overlap with the final prefill chunk, enabling decode to begin in the next iteration.

\subsection{Design Components}

\parab{High-performance heterogeneous execution.}
Efficient XP execution requires preserving CUDA-graph execution while limiting interference between phase-specific attention paths.
\sys captures the linear operators in piecewise graph segments and uses SM partitioning to limit interference in attention-path overlap (\S\ref{sec:engine-execution}).

\parab{Latency modeling and estimation.}
\para's heterogeneous execution makes iteration latency depend on the slowest rank and execution path at its synchronization boundaries.
We develop a latency model that captures these rank- and path-dependent bottlenecks, and construct a profile-fitted estimator
 to ensure TBT-safe co-batching (\S\ref{sec:estimator}).

\begin{figure}[tbp]
    \centering
    \includegraphics[width=1.0\columnwidth]{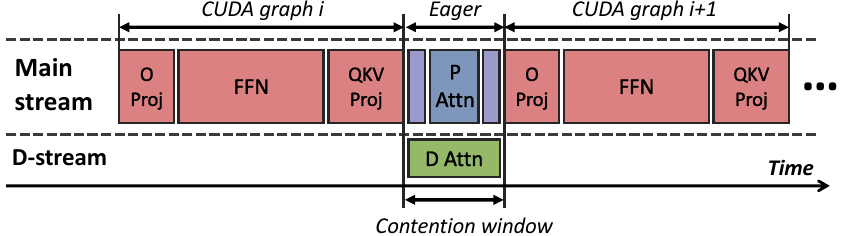}
    \caption{
    Piecewise CUDA-graph integration of \para.
    Shared linear operators remain graph-captured, while phase-specific attention paths execute concurrently on separate CUDA streams in a narrow eager region, confining cross-phase contention to the custom attention operator.
    }
    \label{fig:piecewise-cg-impl}
\end{figure}

\parab{Model-guided dispatch and placement.}
{
As \para's execution latency depends on rank-local load, both prefill dispatch and decode placement affect the TBT-safe prefill budget. Rank imbalance increases the bottleneck load, leaving {bubbles} on other ranks and reducing the admissible TBT-safe prefill work.
Guided by the latency model, \sys maximizes the TBT-safe prefill budget %
through two coupled decisions:
(1) waterfill dispatch assigns prefill tokens to balance the current combined prefill--decode load across ranks; and (2) dual-mode decode placement shapes future resident load, spreading requests to reduce execution imbalance or packing them to preserve KV capacity (\S\ref{sec:dispatcher}).}

\parab{SLO-aware scheduler design.}
Maximizing SLO-compliant goodput requires deciding which prefills should consume the TBT-safe budget. \sys integrates TTFT-aware admission and prioritization with model-guided prefill dispatch and decode placement within the scheduling loop (\S\ref{sec:scheduler}).

\section{\sys: Design}

\subsection{High-Performance Heterogeneous Execution}
\label{sec:engine-execution}

\paragraph{Piecewise graph execution.}
We leverage piecewise CUDA graphs~\cite{vllm2026cudagraphs} to efficiently realize XP's split execution structure.
As illustrated in Figure~\ref{fig:piecewise-cg-impl}, shared linear operators remain within graph-captured segments, while the attention region is replaced by our custom heterogeneous attention operator.
Execution therefore alternates between graph-captured shared-linear segments
and narrow heterogeneous-attention regions.
This decomposition (1) preserves optimized CUDA-graph execution for all shared linear operators, (2) isolates heterogeneous execution to a narrow attention region, which enables coordinated attention overlap.

\paragraph{Coordinated attention overlap.}
Within the custom attention operator, \sys executes the prefill and decode attention paths concurrently on separate CUDA streams. The distributed prefill path, including its \texttt{All-to-All} communication and attention kernel, runs on the main stream, while localized decode attention runs on the D stream.
Concurrent prefill and decode attention can introduce SM contention. \sys therefore uses GreenContext~\cite{nvidia2025greencontexts} to partition SMs within the custom operator. We allocate decode the minimum number of SMs needed to sustain 90\% of its standalone peak HBM bandwidth, leaving the remaining SMs to prefill. %

\subsection{Latency Modeling and Estimation}
\label{sec:estimator}

\begin{figure}[tbp]
    \centering
    \includegraphics[width=0.96\columnwidth]{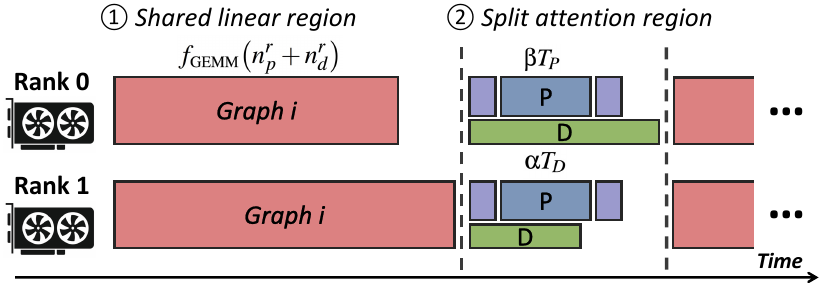}
    \caption{
    Latency model across two logical synchronization boundaries. Rank- and path-level imbalance introduces idle time in shared linear and split attention regions.
    }
    \label{fig:modeling-balancing}
\end{figure}

\parab{Synchronization Boundaries.}
As Figure~\ref{fig:modeling-balancing} shows, \para execution exposes two logical synchronization boundaries per Transformer layer: one after the shared linear region (\raisebox{-0.15ex}{\ding{172}}) and one after the split attention region (\raisebox{-0.15ex}{\ding{173}}). We therefore decompose per-iteration latency across $L$ layers as
\begin{equation}
T_{\mathrm{iter}}
=
L\left(
T_{\scriptscriptstyle \mathrm{GEMM}} + T_{\scriptscriptstyle \mathrm{ATTN}}
\right),
\label{eq:xp_latency}
\end{equation}
where each term is determined by the slowest rank or execution path reaching the corresponding boundary. For compactness, we express component latencies using base latency functions $f_{\scriptscriptstyle \mathrm{GEMM}}$, $f_D$, $f_P$, and $f_{\mathrm{comm}}$.
We detail the functional forms of these components in Appendix~\ref{app:latency-model}.

\parab{Shared linear region.}
Let $n_p^r$ and $n_d^r$ denote the numbers of prefill and decode tokens processed on rank $r$, respectively. Since each rank processes its local prefill and decode tokens together through the shared linear operators,

\begin{equation}
T_{\scriptscriptstyle \mathrm{GEMM}}
=
\max_r f_{\scriptscriptstyle \mathrm{GEMM}}\!\left(n_p^r+n_d^r\right)
=
f_{\scriptscriptstyle \mathrm{GEMM}}\!\bigl(\max_r(n_p^r+n_d^r)\bigr),
\label{eq:linear_boundary}
\end{equation}

where $f_{\scriptscriptstyle \mathrm{GEMM}}(\cdot)$ is a nondecreasing function of the rank-local token count. Thus, the shared-linear boundary is determined by the largest combined token load across ranks.

\parab{Split attention region.}
At the second boundary, prefill follows the distributed SP attention path, while decode attention remains rank-local. Let $T_P$ and $T_D$ denote their standalone latencies. Since the two paths execute concurrently,
\begin{equation}
T_{\scriptscriptstyle \mathrm{ATTN}}
=
\max\left\{
\beta T_P,\;
\alpha T_D
\right\},
\label{eq:attn_boundary}
\end{equation}
where $\alpha,\beta\geq1$ capture the slowdowns caused by contention for the decode and prefill paths, respectively.

For decode, each request remains localized to its resident rank, so the path latency is determined by the slowest rank:
\begin{equation}
T_D
=
\max_r
f_D\!\biggl(
\sum_{i\in\mathcal{D}^r} c_i
\biggr),
\label{eq:decode_attn}
\end{equation}
where $\mathcal{D}^r$ denotes the decode requests resident on rank $r$, and $c_i$ is the context length of request $i$.

For prefill, the distributed path consists of \texttt{All-to-All} communication and attention. Let $\mathcal{P}$ denote the scheduled prefill requests, with processed context length $c_i$ and scheduled chunk size $\ell_i$ for request $i$. We model its latency as
\begin{equation}
T_P
=
f_{\mathrm{comm}}\!\left(\max_r n_p^r\right)
+
f_P\!\left(
\frac{1}{N}
\sum_{i\in\mathcal{P}}
\ell_i\left(c_i+\frac{\ell_i+1}{2}\right)
\right).
\label{eq:prefill_attn}
\end{equation}
The communication term is determined by the largest prefill shard, $\max_r n_p^r$, where $n_p^r$ denotes the number of prefill tokens assigned to rank $r$.
The attention term captures the aggregate prefill attention workload, which Ulysses-SP distributes evenly across the $N$ ranks, independent of the per-rank prefill token assignment.

\parab{Estimator construction.}
We instantiate the model through offline profiling.
We profile the shared linear operators, prefill attention, decode attention, and communication in isolation to obtain $f_{\scriptscriptstyle \mathrm{GEMM}}$, $f_P$, $f_D$, and $f_{\mathrm{comm}}$, and use heterogeneous \para profiles to calibrate the contention factors $\alpha$ and $\beta$.
The resulting estimator closely matches measured iteration latency across diverse configurations (\S\ref{subsec:latency-acc}).

\subsection{Model-Guided Dispatch and Placement}
\label{sec:dispatcher}

\parab{Prefill: waterfill dispatch.}

{
Given the current decode state and TBT SLO, we seek a prefill assignment $\mathbf A_{\mathrm p}=\{n_p^r\}_{r=1}^{N}$ that maximizes the TBT-safe prefill budget $B=\sum_r n_p^r$.
}

The latency model provides key guidance for dispatch.
For the shared linear region (Eq.~\ref{eq:linear_boundary}), latency depends on the largest combined prefill--decode load, $\max_r(n_p^r+n_d^r)$. Since $n_d^r$ is fixed within an iteration, this suggests assigning more prefill tokens to decode-light ranks.
For the split attention region (Eq.~\ref{eq:prefill_attn}), prefill communication depends on $\max_r n_p^r$, while prefill-attention computation is independent of the cross-rank prefill assignment because Ulysses-SP {evenly} redistributes the aggregate attention workload across ranks.
Importantly, our profiling shows that shared-linear latency typically dominates prefill communication latency.%

Together, these observations yield an important dispatch principle: maximizing the TBT-safe prefill budget primarily requires \textit{balancing the combined token load}, $n_p^r+n_d^r$, across ranks.
This reduces the dispatch problem to finding the largest common combined-load target $n_{\mathrm{target}}$ satisfying the TBT SLO, then waterfilling prefill tokens toward this target.

{%
We design a waterfill dispatcher based on this insight.
The dispatcher uses binary search with the latency estimator to find $n_{\mathrm{target}}$.
Under this target, rank $r$ has capacity $\max\{0,n_{\mathrm{target}}-n_d^r\}$.
For a given request, the aggregate capacity determines the TBT-safe budget $B$, capped by its remaining prefill length, and waterfilling these tokens toward $n_{\mathrm{target}}$ yields $\mathbf A_{\mathrm p}$.
With $N$ ranks and a target-level search range of $B$, each invocation takes $O(N\log B)$ time, since each latency-estimator query takes $O(N)$.
}

{%
During scheduling, as shown in Algorithm~\ref{alg:slo-aware-scheduler}, the dispatcher is invoked incrementally across prefill requests. For request $i$, it returns {the largest} TBT-safe chunk $\ell_i$ and corresponding per-rank assignment $\mathbf A_i$, given the existing assignment $\mathbf A_{\mathrm p}$ (Line~6). The scheduler then updates $\mathbf A_{\mathrm p}$ with $\mathbf A_i$ (Line~13). Appendix~\ref{app:dispatch} provides the dispatcher pseudocode.
}

\parab{Decode: dual-mode placement.}
When a prefill completes, \sys places it on a single rank %
for KV residency and decode execution.
The placement aims to maximize the TBT-safe prefill budget in subsequent iterations while preserving sufficient KV capacity for new decode placements.

To maximize the TBT-safe prefill-dispatch budget, the latency model suggests a simple placement principle: place each decode request on the rank with the \textit{smallest aggregate decode context}, $\sum_{i\in\mathcal{D}^r} c_i$.
For the split attention region (Eq.~\ref{eq:decode_attn}), placing a request on the rank with the smallest aggregate decode context, $\sum_{i\in\mathcal D^r} c_i$, minimizes decode imbalance. For the shared linear region (Eq.~\ref{eq:linear_boundary}), a placement increases $n_d^r$ by only one on the selected rank, {while the resulting imbalance can be compensated by subsequent waterfill prefill dispatch.}

KV capacity, however, favors tighter packing. Spreading decode requests can fragment free KV capacity across ranks, limiting future placements.
\sys therefore reconciles these objectives using a KV-utilization threshold $\eta$.
While any feasible rank remains below $\eta$, it uses worst-fit placement, selecting the rank with the smallest aggregate decode context.
If all ranks exceed $\eta$, it switches to best-fit to reduce fragmentation.
Appendix~\ref{app:placement} provides the pseudocode.

\subsection{SLO-Aware Scheduler Design}
\label{sec:scheduler}

Within each XP instance, the scheduler determines which pending prefills should consume the {TBT-safe budget} available under the resident decode workload.
As shown in Algorithm~\ref{alg:slo-aware-scheduler}, it first performs TTFT-aware admission and prioritization (Lines~1--4).
A Moore--Hodgson admission policy~\cite{moore1968} (Appendix~\ref{app:admission}) selects requests that can meet their TTFT deadlines under the {reference budget $B_0$}, with admitted requests considered in EDF order; the remaining requests are ordered by SJF for opportunistic backfill.

Following this order, the scheduler constructs the prefill batch incrementally (Lines~5--13).
{
For each request, it invokes the prefill dispatcher to obtain a TBT-safe chunk and corresponding assignment under the current state.
}
If the chunk completes the request's prefill, the scheduler further invokes decode placement before admitting the transition.
An infeasible placement discards the completing chunk and its tentative assignment and stops further admission. Otherwise, the selected decode rank is reserved and incorporated into the tentative decode state so that subsequent placement decisions account for earlier transitions in the same iteration.
{
The accepted chunk is then added to the prefill batch, and its assignment is accumulated into the cumulative per-rank assignment.
}
Let $R=|\mathcal R|$ be the number of pending prefill requests and $N$ the number of ranks. Admission and ordering require $O(R\log R)$ time, while at most $O(R)$ dispatcher invocations each take $O(N\log B)$. Transition validation takes $O(RN)$ time in the worst case. The overall per-iteration complexity is therefore $O(R\log R+RN\log B)$.

\begin{algorithm}[tb]
\caption{SLO-Aware Scheduler}
\label{alg:slo-aware-scheduler}
\small
\begin{algorithmic}[1]

\Require Pending prefills $\mathcal R$, decode state $\mathcal D$, time $t$, TBT SLO $\tau$
\Ensure Prefill assignment $\mathbf A_{\mathrm p}$,
decode placement $\mathbf A_{\mathrm d}$

\State $\mathcal P\gets\emptyset$;
$\mathbf A_{\mathrm p}\gets\emptyset$;
$\widetilde{\mathcal D}\gets\mathcal D$;
$\mathbf A_{\mathrm d}\gets\emptyset$

\State $B_0\gets\Call{Dispatch}{\tau,\mathcal D,\mathcal P,\mathbf A_{\mathrm p},\emptyset}$
\State $(\mathcal R_{\mathrm{adm}},\mathcal R_{\mathrm{rej}})
    \gets\Call{MooreHodgsonAdmission}{\mathcal R,B_0,t,\tau}$

\State Sort $\mathcal R_{\mathrm{adm}}$ in EDF order;
sort $\mathcal R_{\mathrm{rej}}$ in SJF order

\State \textbf{for each} $i$ in
$\mathcal R_{\mathrm{adm}}$ followed by $\mathcal R_{\mathrm{rej}}$

\IState $(\ell_i,\mathbf A_i)\gets\Call{Dispatch}{\tau,\mathcal D,\mathcal P,\mathbf A_{\mathrm p},i}$
\IState \textbf{if} $\ell_i=0$ \textbf{then break} \Comment{Prefill budget exhausted}

\IState \textbf{if} $\ell_i=m_i$ \textbf{then} \Comment{Prefill completes}
\IIState $g_i\gets\Call{DecodePlacement}{i,\widetilde{\mathcal D}}$
\IIState \textbf{if} $g_i=\textsc{Infeasible}$ \textbf{then}
\Comment{No feasible decode host}
\IIIState \textbf{break}
\IIState $\mathbf A_{\mathrm d}
    \gets\mathbf A_{\mathrm d}\cup\{(i,g_i)\}$;
$\widetilde{\mathcal D}\gets\widetilde{\mathcal D}\cup\{(i,g_i)\}$

\IState $\mathbf A_{\mathrm p}\gets\mathbf A_{\mathrm p}\cup\mathbf A_i$;
$\mathcal P\gets\mathcal P\cup\{(i,\ell_i)\}$

\State \textbf{return} $\mathbf A_{\mathrm p},\mathbf A_{\mathrm d}$

\end{algorithmic}
\end{algorithm}

\section{Implementation}
\label{sec:implementation}

We implement \sys on vLLM v0.15.1~\cite{kwon2023efficient} in approximately 6K lines of code. We reuse vLLM's continuous batching, PagedAttention, and FlashAttention~\cite{shah2024flashattention3} backends, and adapt the Ulysses-SP implementation from the ArcticInference plugin~\cite{hidayetoglu2026shif} for \para.

Beyond optimizations mentioned in \S\ref{sec:engine-execution}, \sys further implements two optimizations for phase-specific KV layouts.
First, KV-cache aliasing allows head-sharded prefill and localized decode to share the same physical PagedAttention blocks, while separate block tables and address mappings expose the corresponding views to their attention kernels. This avoids KV duplication and separate KV pools.
 Second, during a prefill-to-decode transition, \sys preselects the decode rank before the final prefill chunk completes and pipelines KV transfer layer by layer. After layer $i$ finishes attention, its KV shard is transferred asynchronously while layer $i+1$ executes its graph-captured GEMM segment, overlapping {transfer} with the prefill execution.

\section{Evaluation}
\label{sec:evaluation}

We evaluate \sys to answer the following questions:

\begin{itemize}
    \item How does \sys compare with state-of-the-art systems across different
    models? (\S\ref{sec:eval-e2e})

    \item How do the key design components contribute to \sys's performance gains?
    (\S\ref{sec:eval-abl})

    \item How do workload characteristics and SLO {tightness} affect \sys's
    performance? (\S\ref{sec:eval-sen})

    \item How accurate is \sys's latency estimator, and what runtime overhead
    does \sys introduce? (\S\ref{sec:eval-component})
\end{itemize}

\subsection{Setup}
\label{subsec:eval-setup}

\begin{figure*}[tb]
    \centering
    \includegraphics[width=0.95\textwidth]{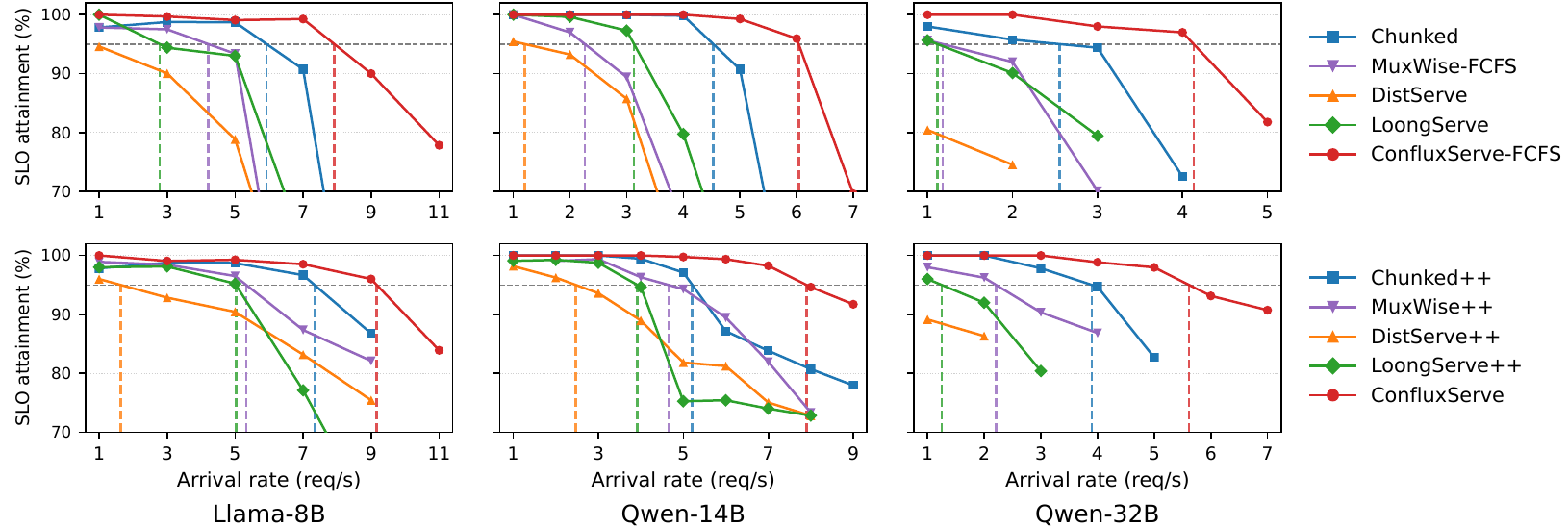}
    \caption{
    SLO attainment on Mooncake-Toolagent, with FCFS scheduling (top) and enhanced scheduling configurations (bottom). Vertical lines indicate the maximum arrival rate a system sustains while meeting an SLO attainment of over 95\%.
    }
    \label{fig:e2e-slo-attainment}
\end{figure*}

\parab{Models.}
We evaluate \sys on three widely used open-source models spanning different sizes: Llama-3.1-8B~\cite{grattafiori2024llama3}, Qwen3-14B, and Qwen3-32B~\cite{yang2025qwen3}, abbreviated as Llama-8B, Qwen-14B, and Qwen-32B. Following {prior works~\cite{zhang2026jitserve,chen2026towards, du2025ecoserveenablingcosteffectivellm}}, we use TTFT/TBT SLOs of 1s/50ms for Llama-8B and 2s/100ms for Qwen-14B and Qwen-32B.

\begin{table}[tb]
  \centering
  \small
  \renewcommand{\arraystretch}{0.92}
  \setlength{\tabcolsep}{4pt}
  \begin{tabular}{l *{4}{>{\raggedleft\arraybackslash}p{28.5pt}}}
    \toprule
    \noalign{\vskip -1pt}
    Workload & \multicolumn{2}{c}{Prompt (P50/P95)}
             & \multicolumn{2}{c}{Output (P50/P95)} \\[-1pt]
    \midrule
    \noalign{\vskip -0.5pt}
    Mooncake-Toolagent & 6{,}346  & 26{,}114 & 30  & 600 \\
    Mooncake-Synthetic & 11{,}587 & 49{,}009 & 69  & 519 \\
    QwenA              & 1{,}046  & 8{,}809  & 375 & 1{,}032 \\[-0.5pt]
    \bottomrule
  \end{tabular}
    \caption{Workload prompt and output length statistics.}
    \label{tab:workloads}
\end{table}

\parab{Metrics.}
We measure request-level SLO attainment ~\cite{agrawal2024etalon,goel2025niyama}: for the $i$-th output token, %
the deadline relative to request arrival is $\mathrm{TTFT}_{\mathrm{SLO}}+(i-1)\cdot\mathrm{TBT}_{\mathrm{SLO}}$. A request is SLO-compliant only if every output token meets its corresponding deadline, jointly capturing prefill and decode performance.
Our primary metric is \emph{goodput}: the maximum sustainable request rate at which at least 95\% of requests are SLO-compliant.

\parab{Testbed.}
We conduct experiments on NVIDIA H100 SXM5 80GB GPUs connected via NVLink. We use four GPUs for Llama-8B and Qwen-14B and eight GPUs for Qwen-32B.
Each configuration forms one XP instance. Due to the available testbed scale, we evaluate a single instance and therefore do not exercise global routing.
All key design choices in \para and \sys are evaluated.
All compared systems use the same number of GPUs for each model.
Our software stack uses CUDA 12.8, NCCL 2.27.5, and PyTorch 2.9.1. We use vLLM v0.15.1~\cite{kwon2023efficient} and SGLang v0.5.9~\cite{zheng2024sglang} for the corresponding baselines. For LoongServe, we use its open-source implementation and update its CUDA, NCCL, and PyTorch dependencies to match our testbed.

\parab{Workloads.}
We evaluate three representative workloads derived from the Mooncake~\cite{qin2025mooncake} and Qwen~\cite{wang2025kvcachewild} traces: Mooncake-Toolagent, Mooncake-Synthetic, and QwenA.
As summarized in Table~\ref{tab:workloads}, Mooncake-Toolagent has long-tailed prompts and short outputs, representative of tool-and-agent workloads. Mooncake-Synthetic has substantially longer prompts, while QwenA has shorter prompts and much longer outputs.
We use Mooncake-Toolagent for the main end-to-end evaluation and the other two for sensitivity analysis.

We subsample each trace
and uniformly scale inter-arrival times to obtain the target arrival rate. We cap prompt and output lengths at each model's maximum context length.

\parab{Baselines.}
We compare \sys against four state-of-the-art systems spanning the prefill--decode execution architectures in Table~\ref{tab:pd-taxonomy}: Chunked Prefill~\cite{agrawal2024taming}, DistServe~\cite{zhong2024distserve}, LoongServe~\cite{wu2024loongserve}, and MuxWise~\cite{chen2026towards}. For each baseline, we retain the parallelism strategy (\textit{i.e.,} TP or SP) prescribed by its original design and implementation.

To separate execution benefits from request-scheduling effects, we evaluate two
settings. First, we normalize all systems to FCFS, disabling MuxWise's preemption mechanism (MuxWise-FCFS), to isolate the underlying prefill--decode execution architecture.
We then enhance each baseline with the same SRPT-style policy, prioritizing requests with the fewest remaining prefill tokens, and denote these variants with a ``++'' suffix.
MuxWise++ outperforms the original MuxWise scheduler under the evaluated workloads.

\begin{itemize}
\item \textbf{Chunked Prefill (co-batching).}
We use vLLM's chunked-prefill execution with TP4 for Llama-8B and Qwen-14B and TP8 for Qwen-32B. %
For each model, workload, and TBT target, we sweep the token budget and report the best-performing configuration.

\item \textbf{MuxWise (spatial multiplexing).}
We use the MuxWise implementation in SGLang with TP4 for Llama-8B and Qwen-14B and TP8 for Qwen-32B. %
{We retain its TBT-aware SM allocation, assigning just enough SMs to decode to meet the TBT SLO and the remaining SMs to prefill.}

\item \textbf{DistServe (static isolation).}
We use vLLM's disaggregated serving with NCCL KV connector.
{We sweep feasible prefill--decode GPU allocations} and use the best-performing 1:1 split: We use TP2 per phase for Llama-8B and Qwen-14B, and TP4 per phase for Qwen-32B.

\item \textbf{LoongServe (elastic isolation).}
We use the original {Loong-Serve} implementation with base TP1 and maximum SP4 for Llama-8B and Qwen-14B, and base TP2 and maximum SP4 for Qwen-32B, where TP1 leaves insufficient KV-cache headroom after loading the model weights.
We retain its execution architecture and enable piecewise CUDA graphs to align launch overhead with the other systems.

\end{itemize}
For \sys, XP instances use base TP1 with SP4 for Llama-8B and Qwen-14B, and base TP2 with SP4 for Qwen-32B, following the same rationale as LoongServe.

\subsection{End-to-End Performance}
\label{sec:eval-e2e}

Figure~\ref{fig:e2e-slo-attainment} reports request-level SLO attainment on Mooncake-Toolagent across the three models. The top {row} compares all systems under FCFS scheduling, while the bottom compares the scheduling-enhanced baselines with full \sys.

{
We first examine FCFS to understand the execution benefit of \para.
Most systems achieve high SLO attainment at low to moderate load but degrade at markedly different rates as the arrival rate increases.
}
DistServe and LoongServe degrade earliest because phase isolation prevents queued prefills from using compute capacity left underutilized by decode: DistServe uses a static prefill--decode partition, while LoongServe dynamically reallocates resources across phases but still executes them separately.
MuxWise improves cross-phase utilization through spatial multiplexing, but separate prefill and decode kernels forgo
{co-batching's weight reuse}
and introduce cross-kernel resource contention.
Chunked Prefill performs better by co-batching the two phases, but its uniform parallelism configuration
{consumes decode-latency headroom and constrains the TBT-safe prefill budget.}
In contrast, \para combines co-batching with phase-specific parallelism, allowing \sys-FCFS to maintain high SLO attainment at substantially higher loads. It sustains approximately 7.9, 6.0, and 4.1 requests/s on Llama-8B, Qwen-14B, and Qwen-32B, outperforming the strongest baseline by $1.33\times$, $1.34\times$, and $1.62\times$, respectively.

The scheduling-enhanced setting shows a similar trend,
and \sys retains a clear advantage as load increases. With its SLO-aware scheduler, \sys sustains approximately 9.2, 7.9, and 5.6 requests/s on Llama-8B, Qwen-14B, and Qwen-32B, respectively, exceeding the strongest scheduling-enhanced baseline by $1.26\times$, $1.52\times$, and $1.44\times$. Relative to \sys-FCFS, the SLO-aware scheduler further improves goodput by $1.16\times$, $1.32\times$, and $1.37\times$, respectively, demonstrating the
 additional
benefit of our scheduler design.

\subsection{Ablation Study}
\label{sec:eval-abl}

We evaluate how key design components contribute to \sys. Unless otherwise specified, experiments use Qwen-14B on a four-GPU \inst (TP1--SP4).

\begin{figure}[tbp]
    \centering
    \includegraphics[width=\columnwidth]{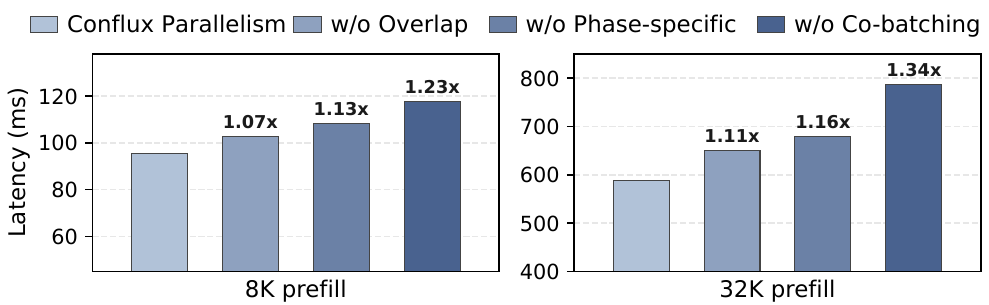}
    \caption{Component-wise benefit of \para with 8K- and 32K-token prefills. Each configuration progressively removes attention overlap, phase-specific parallelism, and prefill--decode co-batching.  Values denote latency normalized to full \para.
    }
    \label{fig:config-compare}
\end{figure}

\begin{figure}[tbp]
    \centering
    \includegraphics[width=1.0\columnwidth]{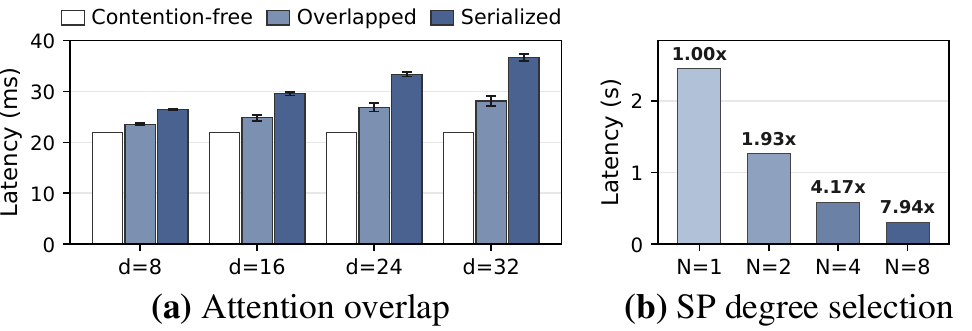}
    \caption{
Overlapping and parallelism design in \para. (a)
\overlap introduces up to 1.3$\times$ speedup with highly stable performance.  (b) parallelism configuration for XP instance achieves near-linear scaling.
}
    \label{fig:overlap-scaling}
\end{figure}

\parab{\para performance breakdown.}
To {isolate} the component-wise benefits of \para, we {maintain} 24 decode requests per GPU, each with 8K-token contexts under a 100\,ms TBT SLO, and measure the completion time of an
injected 8K- or 32K-token prefill request.
Figure~\ref{fig:config-compare} progressively removes attention overlap, phase-specific parallelism, and prefill--decode co-batching from \para.
For the 8K prefill, disabling attention overlap increases latency to $1.07\times$ that of \para. Further removing phase-specific parallelism increases latency to $1.13\times$, while additionally removing co-batching raises it to $1.23\times$. The impact becomes more pronounced for the 32K prefill: {the
slowdowns increase to $1.11\times$, $1.16\times$, and $1.34\times$, respectively.}
The larger gap reflects the cumulative cost of these execution inefficiencies across multiple prefill chunks.

\parab{Effectiveness of coordinated \overlap.}
{
Figure~\ref{fig:overlap-scaling}(a) evaluates attention overlap by co-executing one 8K-token prefill while increasing the decode load from 8 to 32 requests per GPU.
}
We compare \para's overlapped attention execution against serialized execution and an idealized contention-free lower-bound derived from isolated component latencies.
As the decode load increases, the opportunity for overlap grows: at 32 decode requests per GPU, \para reduces attention-region latency from 36.5\,ms to 28.0\,ms, a $1.3\times$ speedup over serialized execution, closing 58\% of the gap to the idealized contention-free lower bound.
{
Moreover, \overlap exhibits a coefficient of variation below 3.5\% across 20 runs for every configuration, indicating stable interference under coordinated overlap.
}

\parab{Parallelism configuration for \inst.}
Figure~\ref{fig:overlap-scaling}(b) validates our choice of using {a fixed maximum intra-node SP degree for each \inst.}
We fix the base TP degree to 1 and scale a 32K-token prefill from SP1 to SP8,
while keeping the per-GPU decode load at 24 requests with 8K-token contexts under a 100\,ms TBT SLO.
Increasing the SP degree from 1 to 2, 4, and 8 reduces {prefill completion latency by $1.93\times$, $4.17\times$, and $7.94\times$, respectively.}
Although higher SP introduces additional communication, \para overlaps the distributed prefill path with local decode attention, effectively hiding much of this overhead and enabling near-linear prefill scaling when co-executed with decode.

\begin{figure}[tb]
    \centering
    \includegraphics[width=0.96\columnwidth]{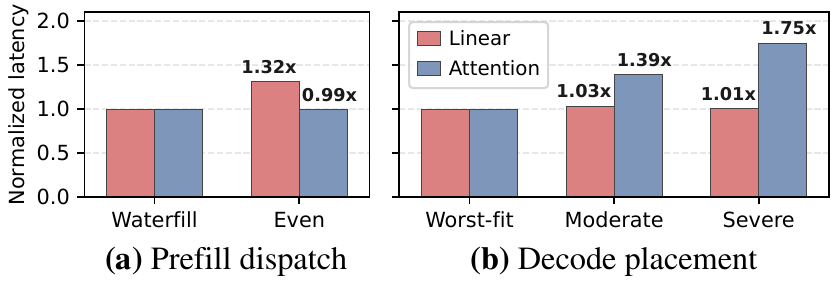}
    \caption{Effectiveness of model-guided prefill dispatch and decode placement. Latencies are normalized to our solution.
    }
    \label{fig:dispatch-boundaries}
\end{figure}

\parab{Effectiveness of prefill dispatch.}
{
Figure~\ref{fig:dispatch-boundaries}(a) evaluates waterfill under a skewed four-GPU decode load: one rank hosts 128 decode requests with 2K-token contexts, while the other three have no decodes. We co-execute a 1K-token prefill and compare against even dispatch.
}
{By assigning fewer prefill tokens to the decode-heavy rank, waterfill balances the combined load, while even dispatch incurs $1.32\times$ the shared-linear latency.}
Meanwhile, split-attention latency remains nearly unchanged ($0.99\times$), as Ulysses-SP redistributes the prefill workload across ranks.
Together, waterfill reduces the critical-rank bottleneck without penalizing prefill attention.

\parab{Effectiveness of decode placement.}
Figure~\ref{fig:dispatch-boundaries}(b) evaluates
 the benefit of
worst-fit decode placement {when sufficient KV headroom is available.} We fix 64 decode requests, each with an 8K-token context, and co-execute a 1K-token prefill.
We evaluate three decode placements: \textit{worst-fit} $[16,16,16,16]$, \textit{moderately skewed} $[8,8,24,24]$, and \textit{severely skewed} $[0,0,32,32]$.
Shared-linear latency remains nearly unchanged ($1.03\times$ and $1.01\times$) because waterfill prefill dispatch compensates for the uneven decode load. In contrast, split-attention latency
increases to $1.39\times$ and $1.75\times$ under moderate and severe skew, respectively,
as the most heavily loaded rank determines the critical path.
This validates our choice of worst-fit placement to balance decode load across ranks.

\label{sec:eval-parallelism-degree}

\subsection{Sensitivity Study}
\label{sec:eval-sen}

\begin{figure}[tb]
    \centering
    \includegraphics[width=0.94\columnwidth]{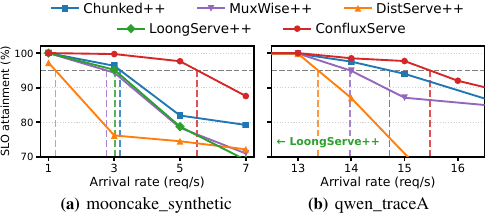}
    \caption{
    \sys on Qwen-14B with Mooncake-Synthetic and QwenA  (other setups follow \S\ref{sec:eval-e2e}).
    }
    \label{fig:dataset-slo-attainment}
\end{figure}

\begin{figure}[tb]
    \centering
    \includegraphics[width=0.94\columnwidth]{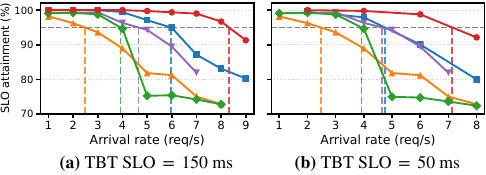}
    \caption{
\sys performance on Qwen-14B with different TBT SLOs (other setups follow \S\ref{sec:eval-e2e}).
}
    \label{fig:var-slo-attainment}
\end{figure}

\parab{Workload Sensitivity.}
We evaluate Qwen-14B on two additional workloads, Mooncake-Synthetic and QwenA, using the same setup and SLO targets in \S\ref{sec:eval-e2e}.
As summarized in Table~\ref{tab:workloads}, Mooncake-Synthetic is more prefill-heavy, with longer prompts and similarly short outputs, whereas QwenA is more decode-heavy with shorter prompts and longer outputs.
{Figure~\ref{fig:dataset-slo-attainment} shows that \sys sustains the highest load on both workloads.
On Mooncake-Synthetic, \sys sustains 5.5 requests/s, $1.7\times$ {that of} the strongest baseline, showing that its advantage persists under substantially longer prompts.
On QwenA, \sys sustains 15.6 requests/s, $1.1\times$ {that of} the strongest baseline. The smaller gain is expected as shorter prompts %
reduce the prefill work {available} for exploiting prefill--decode resource complementarity. \sys still benefits from localized decode execution and avoids the uniform-parallelism tax. %
}

\parab{TBT Sensitivity.}
We vary the default 100\,ms TBT SLO to {150\,ms and 50\,ms} on Qwen-14B.
As shown in Figure~\ref{fig:var-slo-attainment}, \sys sustains {8.3 requests/s at 150\,ms and 7.2 requests/s at 50\,ms}, outperforming the strongest baseline by $1.4\times$ and $1.5\times$, respectively.
{
Relaxing the TBT SLO provides additional headroom for prefill co-execution, increasing the sustainable rate of \sys and Chunked++. The other baselines gain little since they cannot effectively use this headroom for prefill progress.}
Conversely, tightening the TBT SLO reduces the TBT-safe prefill budget, slightly reducing the
{sustainable rate}
of both \sys and Chunked++.
\sys nevertheless retains its advantage, as its phase-specific parallelism preserves more decode-latency headroom than Chunked++.

\subsection{Component Accuracy and Overheads}
\label{sec:eval-component}

\paragraph{Latency-Estimator Accuracy}
\label{subsec:latency-acc}

{
We evaluate the latency estimator using mean absolute percentage error (MAPE) on 200 randomly sampled heterogeneous batches per model, spanning prefill chunks of 32--10K tokens, decode batches of 1--128, and context lengths of 1K--16K tokens.
}
{As shown in Table~\ref{tab:latency-estimator}, the estimator achieves 4.1--4.8\% MAPE
across the three models, showing that it accurately captures \para execution latency across model sizes.

\begin{table}[tb]
    \centering
    \small
    \setlength{\tabcolsep}{5pt}
    \begin{tabular}{l ccc}
        \toprule
        \noalign{\vskip -2pt}
         & \makecell{Llama-8B} & \makecell{Qwen-14B} & \makecell{Qwen-32B} \\[-2pt]
        \midrule
        \noalign{\vskip -1pt}
        MAPE & 4.4\% & 4.8\% & 4.1\% \\[-1pt]
        \bottomrule
    \end{tabular}

    \caption{
    Latency-estimation accuracy across models.
    }
    \label{tab:latency-estimator}
\end{table}

\paragraph{Scheduler Overhead}
We measure \sys's scheduling latency with different pending queue depths. As Table~\ref{tab:scheduler-overhead} shows, scheduling overhead remains below 1.05\,ms across all tested queue depths. Since scheduling runs asynchronously with GPU execution, most of this cost can be hidden. %

\section{Related Work}
\label{sec:related}
We discussed prefill--decode execution designs in \S\ref{subsec:existing-designs}; we now
discuss prior work related to \sys in other areas.

\begin{table}[tb]
    \centering
    \small
    \setlength{\tabcolsep}{4pt}
    \begin{tabular}{l cccc}
        \toprule
        \noalign{\vskip -1pt}
        Queue depth & 10 & 50 & 100 & 500 \\[-1pt]
        \midrule
        \noalign{\vskip -1pt}
        Time (ms) & $0.36\!\pm\!0.05$ & $0.47\!\pm\!0.05$ & $0.56\!\pm\!0.05$ & $1.04\!\pm\!0.05$ \\[-1pt]
        \bottomrule
    \end{tabular}

    \caption{
    \sys scheduling latency. Results report mean $\pm$ standard deviation over 100 scheduling steps.
    }
    \label{tab:scheduler-overhead}
\end{table}

\parab{SLO-Aware LLM Scheduling.}
A large body of work studies SLO-aware scheduling to improve goodput under TTFT and TBT SLOs~\cite{fastserve,andes,zhang2026jitserve,chen2025slos,qlm,adaserve,adnan2026cascade}. \sys shares this objective, but its scheduling design addresses the new challenges introduced by \para's heterogeneous execution: accurately modeling its latency, dispatching prefills across ranks, and placing decode requests to realize its execution-efficiency gains as end-to-end goodput.

\parab{Dynamic Parallelism.}
Shift Parallelism~\cite{hidayetoglu2026shif} and Seesaw~\cite{su2025seesaw} dynamically adapt model-wide parallelism \textit{over time}, selecting a configuration better suited to the currently dominant workload or phase. However, each execution interval still uses a single model-wide configuration, forcing co-executed prefill and decode to share the same parallelism. They therefore remain subject to the utilization--efficiency dilemma.

\parab{Operator-level Heterogeneity}
Prior systems exploit operator-level heterogeneity for different objectives. Helix~\cite{bhatia2025helix} decouples attention and FFN sharding to accelerate ultra-long-context decoding, while Hetis~\cite{mo2025hetis} maps dense computation and decode attention differently to better utilize heterogeneous GPUs. POD-Attention~\cite{kamath2025pod} co-executes prefill and decode attention through a fused, SM-aware kernel under a fixed parallelism configuration. These systems do not support co-executed prefill and decode with different inter-GPU layouts. Orca~\cite{yu2022orca} co-batches requests through compatible operators while retaining request-specific attention, accommodating heterogeneous sequence lengths under a common parallelism configuration. In contrast, \sys enables co-executed prefill and decode to retain \textit{different parallelism configurations}. This parallelism heterogeneity further introduces cross-phase KV relocation and rank-aware prefill dispatch and decode placement, which \sys addresses.
\par

\section{Conclusion}

We identify a utilization--efficiency dilemma in existing prefill-decode execution designs. To resolve it, we propose \para, a parallel execution abstraction that co-executes prefill and decode with {different} parallelism within the {same} batch. \para exploits operator-level Transformer structure to construct a {heterogeneous} execution path with different batching and parallelism strategies across operators, combining the best of co-batching and phase-specific execution. Building on \para, we design \sys, an SLO-aware serving system that maximizes cluster goodput under TTFT and TBT constraints.

\bibliographystyle{ACM-Reference-Format}
\bibliography{references}

\clearpage

\appendix

\section{Parallelism Analysis}
\label{app:parallel}

\begin{table*}[tb]
\centering
\small
\renewcommand{\arraystretch}{1.45}
\begin{tabular}{lcccc}
\toprule
\textbf{Parallelism}
& \textbf{Partition}
& \textbf{Communication}
& \textbf{Per-Rank Communication Volume}
& \textbf{Scaling in $P$} \\
\midrule

\textbf{TP}
& Model dimensions / heads
& $2\times$\texttt{All-Reduce}
& $4\left(\dfrac{P-1}{P}\right)BSH$
& $\Theta(1)$ \\

\addlinespace

\textbf{Ring-SP}
& Sequence
& P2P KV circulation
& $2\left(\dfrac{P-1}{P}\right)\dfrac{B(C+S)H}{G}$
& $\Theta(1)$ \\

\addlinespace

\textbf{Ulysses-SP}
& Sequence $\leftrightarrow$ heads
& $2\times$\texttt{All-to-All}
& $2\left(\dfrac{P-1}{P^2}\right)BSH
\left(1+\dfrac{1}{G}\right)$
& $\Theta(1/P)$ \\

\bottomrule
\end{tabular}
\caption{
Per-rank communication of one Transformer block. $P$, $B$, $S$, $C$, and $H$ denote the parallel degree, batch size, current prefill chunk length, historical context length, and hidden dimension, respectively; $G$ is the query-to-KV head ratio, with $G=1$ corresponding to multi-head attention. TP communicates activations associated with the current chunk through two \texttt{All-Reduce} operations. Standard Ring-SP circulates sequence-partitioned K/V states from both the historical context and the current chunk, whereas Ulysses-SP communicates only current-chunk Q/K/V activations and attention outputs. Communication volumes count transmitted tensor elements per rank and exclude local transfers. The TP expression assumes a ring-based \texttt{All-Reduce}, and the Ring-SP expression assumes standard dense pass-KV Ring Attention. Scaling in $P$ holds $B$, $S$, $C$, $H$, and $G$ fixed.
}
\label{tab:parallelism-analysis}
\end{table*}

Table~\ref{tab:parallelism-analysis} compares the per-rank communication costs of TP, Ring-SP, and Ulysses-SP for chunked prefill. Let $P$ denote the parallel degree, $B$ the batch size, $S$ the current prefill chunk length, $C$ the historical context length, $H$ the hidden dimension, and $G$ the query-to-KV head ratio. All three strategies have the same leading-order per-rank computation, $\Theta((BSH^2 + BS(C+S)H)/P)$, so their primary asymptotic difference lies in communication.

\paragraph{Ring-SP versus Ulysses-SP}
The key distinction between the two SP strategies is how they communicate and retain KV state. Standard Ring-SP~\cite{liu2023ring} keeps KV sequence-partitioned and circulates remote K/V blocks across ranks. During chunked prefill, each rank must therefore receive remote K/V states covering both the historical context and the current chunk. Its per-rank communication consequently depends on $C+S$ and remains $\Theta(B(C+S)H/G)$ as $P$ increases. Moreover, this historical-KV communication is incurred again for each subsequent prefill chunk.

Ulysses-SP~\cite{jacobs2023deepspeedulyssesoptimizationsenabling} instead uses an \texttt{All-to-All} to transpose the current-chunk Q/K/V activations from sequence partitioning to head partitioning. Each rank then computes its assigned attention heads for all current-chunk queries using the cached history and current-chunk K/V, after which a second \texttt{All-to-All} restores sequence-partitioned outputs. Historical KV remains head-partitioned and can be reused directly by subsequent chunks, so only activations associated with the current chunk are communicated. As a result, its per-rank communication is independent of $C$ and decreases as $\Theta(BSH(1+1/G)/P)$ with increasing $P$.

\paragraph{Practical trade-offs.}
Communication volume alone does not capture all differences between the two SP strategies. Ring-SP pipelines P2P KV circulation with blockwise attention computation, allowing communication to be hidden when sufficient attention work is available~\cite{liu2023ring}. However, splitting attention into sequence blocks can reduce fused-kernel efficiency~\cite{fang2024usp}, and vanilla Ring-SP with contiguous sequence partitions suffers from substantial load imbalance under causal attention because different ranks cover different portions of the triangular attention matrix~\cite{brandon2023striped,fang2024usp}. Variants such as Striped Attention and load-balanced Ring Attention mitigate this imbalance by redistributing sequence positions across ranks~\cite{brandon2023striped,fang2024usp}. These optimizations do not, however, remove the need to communicate remote KV state in standard pass-KV Ring execution.

Ulysses-SP preserves complete attention computation for each assigned head and can therefore directly use efficient fused attention kernels such as FlashAttention~\cite{jacobs2023deepspeedulyssesoptimizationsenabling,dao2022flashattention}. Its two \texttt{All-to-All} collectives, however, introduce global synchronization points around attention and can become exposed communication overhead when the interconnect is slow or the local attention workload is small. Standard Ulysses also requires the attention heads to be partitionable across the participating ranks, limiting the usable SP degree by the available head count; this constraint becomes more restrictive for GQA and MQA models with fewer KV heads~\cite{fang2024usp}. Hybrid Ulysses--Ring designs have been proposed in part to relax this head-count constraint and adapt communication to hierarchical network topologies~\cite{fang2024usp}.

\paragraph{Comparison with TP}
For reference, standard TP~\cite{shoeybi2020megatronlm} communicates current-chunk activations through the collective operations associated with the attention output projection and FFN. Under a ring-based \texttt{All-Reduce}, its per-rank communication remains $\Theta(BSH)$ as $P$ increases. Like Ulysses-SP, TP communication is independent of the historical context length; unlike Ulysses-SP, however, increasing the parallel degree does not asymptotically reduce the per-rank communication volume.

\paragraph{Effect of grouped-query attention.}
A larger query-to-KV head ratio $G$, as used by grouped-query attention (GQA)~\cite{ainslie2023gqa}, reduces K/V communication in both SP designs. For Ring-SP, the communicated state consists of K/V blocks, so its communication decreases proportionally with $1/G$. For Ulysses-SP, only the K/V portion of the first \texttt{All-to-All} decreases with $G$, while query and output communication remain unchanged. The leading-order attention computation is unaffected because the number of query heads does not change.

Overall, Ring-SP and Ulysses-SP expose complementary trade-offs. Ring-SP enables fine-grained communication--computation overlap and does not fundamentally require head-wise partitioning, but standard execution repeatedly communicates historical KV state, retains constant per-rank communication as $P$ grows, and requires additional mechanisms to avoid causal load imbalance. Ulysses-SP avoids historical-KV communication during chunked prefill and reduces per-rank communication with increasing $P$, while preserving efficient whole-head attention computation, at the cost of global \texttt{All-to-All} synchronization and head-count constraints. For the intra-node parallel degrees considered in \sys, these properties favor Ulysses-SP for distributed prefill.

\section{Operator Analysis}
\label{app:op-analysis}

\begin{table*}[tb]
\centering
\small
\renewcommand{\arraystretch}{1.5}
\begin{tabular}{lcccc}
\toprule
\textbf{Operator}
& \textbf{Compute}
& \textbf{Dominant Tensor Traffic}
& \textbf{Arithmetic Intensity}
& \textbf{vs. Larger $B$} \\
\midrule

\textbf{Prefill Dense Linear}
& $\Theta(BSH^2)$
& $\Theta(H^2 + BSH)$
& $\Theta\!\left(\dfrac{1}{1/H + 1/(BS)}\right)$
& Increases \\

\addlinespace

\textbf{Prefill Attention}
& $\Theta(BSLH)$
& $\Theta(BSH + BLH/G)$
& $\Theta\!\left(\dfrac{1}{1/L + 1/(GS)}\right)$
& Unchanged \\

\addlinespace

\textbf{Decode Dense Linear}
& $\Theta(BH^2)$
& $\Theta(H^2 + BH)$
& $\Theta\!\left(\dfrac{1}{1/H + 1/B}\right)$
& Increases \\

\addlinespace

\textbf{Decode Attention}
& $\Theta(BLH)$
& $\Theta(BH + BLH/G)$
& $\Theta\!\left(\dfrac{1}{1/L + 1/G}\right)$
& Unchanged \\

\bottomrule
\end{tabular}
\caption{Asymptotic compute, dominant tensor traffic, and arithmetic intensity of major LLM operators. $B$ is the number of independent requests, $S$ the number of active prefill tokens per request, $L$ the average number of KV positions attended by each active query, $H=n_qd$ the query hidden width, and $G=n_q/n_{kv}$ the query-to-KV head ratio. Dense-linear rows represent an $H\times H$ projection; QKV, output-projection, and FFN dimensions change only constant factors when they scale with $H$. For attention, we assume a fused exact-attention implementation~\cite{dao2022flashattention} that avoids materializing the attention matrix in HBM; the tensor-traffic expressions capture the Q/K/V/output tensors and abstract away implementation-dependent tiling and reloads. Fixed constants and bytes per element are omitted.}
\label{tab:op-breakdown}
\end{table*}

Table~\ref{tab:op-breakdown} analyzes how cross-request batching affects the two dominant operator classes in LLM inference. We consider independent requests and omit constant factors and element sizes. For attention, we assume a fused exact-attention implementation~\cite{dao2022flashattention} that avoids materializing the attention matrix in HBM.

\paragraph{Dense linear operators.}
Consider an $H\times H$ linear transformation applied to $T$ active tokens. Its computation is $\Theta(TH^2)$, while its dominant tensor traffic is $\Theta(H^2+TH)$: the $H^2$ model weights are shared across all tokens, while activation traffic grows with $T$. Its arithmetic intensity is therefore $\Theta(1/(1/H+1/T))$ and increases with the number of co-batched tokens. For prefill, $T=BS$; for decode, $T=B$. Co-batching requests therefore amortizes model-weight traffic over a larger token batch, an effect also observed in prior operator-level analyses of LLM serving~\cite{zhu2025nanoflow}.

Importantly, this benefit depends only on sharing the same weight matrix and feature dimension. Tokens from different requests, or from prefill and decode, can therefore be concatenated along the token dimension and processed by the same GEMM. In \para, co-batching the two phases increases the effective $T$ seen by the shared QKV projection, output projection, and FFN operators without requiring them to use the same attention parallelism.

\paragraph{Attention.}
Attention exhibits a different batching behavior because each request attends to its own KV state. For $B$ independent requests with $S$ active queries per request and average attended KV length $L$, attention performs $\Theta(BSLH)$ computation and accesses $\Theta(BSH+BLH/G)$ query/output and KV tensors. Its arithmetic intensity is therefore $\Theta(1/(1/L+1/(GS)))$. Unlike dense linear operators, the factor $B$ cancels: adding requests increases attention computation and request-specific KV traffic proportionally. For decode, where each request contributes one active query, this reduces to $\Theta(1/(1/L+1/G))$.

Thus, increasing the cross-request batch size does not asymptotically improve attention arithmetic intensity. This does not imply that attention receives no practical benefit from batching: larger batches can improve kernel occupancy or amortize launch overhead. Rather, attention lacks the model-weight amortization available to dense linear operators because its dominant KV state is request-specific. Prior serving systems similarly exploit this operator-level distinction through selective batching~\cite{yu2022orca}, while specialized kernels such as POD-Attention exploit the complementary resource demands of prefill and decode attention through concurrent execution rather than shared model-weight reuse~\cite{kamath2025pod}.

\paragraph{Implication for \para.}
This asymmetry motivates \para's operator-level execution structure. Co-batching prefill and decode through the shared linear operators captures the arithmetic-intensity benefit of amortizing model-weight traffic, whereas separating their attention paths gives up substantially less cross-request batching benefit while allowing each phase to retain its own attention parallelism. The separated paths can additionally execute concurrently to exploit the complementary resource demands of prefill and decode attention.

\section{\para over a Base TP Configuration}
\label{app:tp-sp}

\begin{figure*}[tb]
    \centering
    \includegraphics[width=2.0\columnwidth]{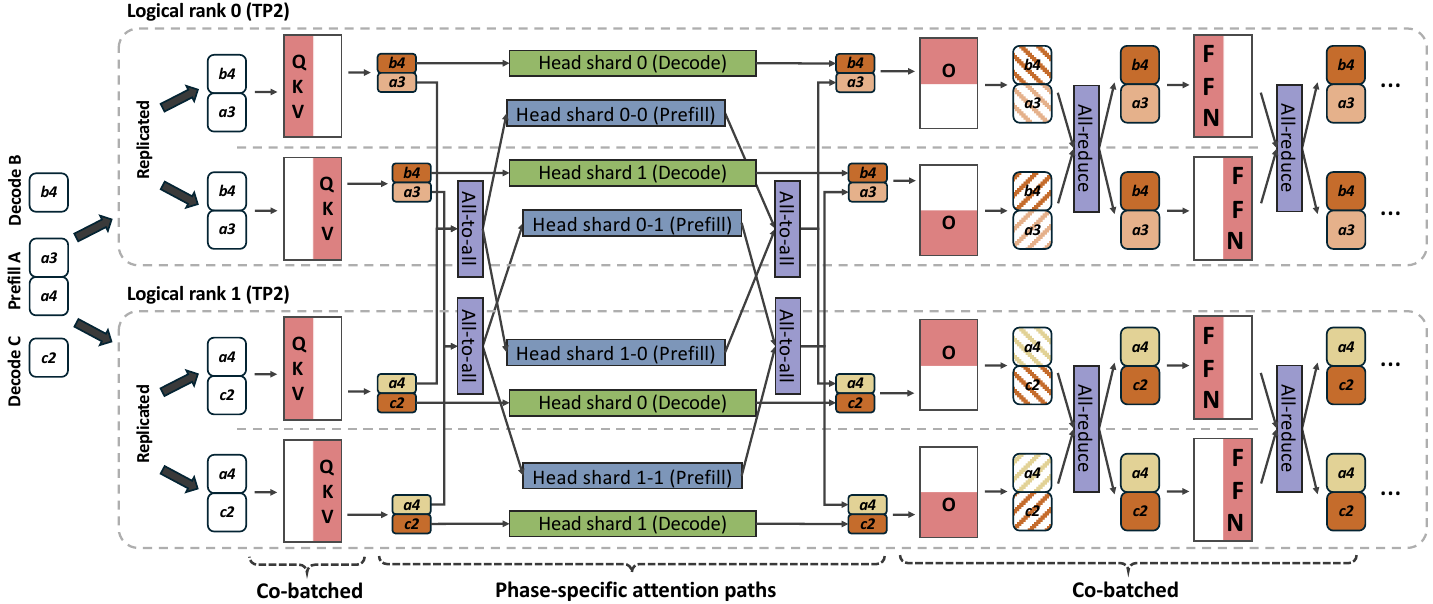}
\caption{
\para composed with a base TP2 configuration across two logical ranks. Each logical rank consists of two TP ranks that share the same rank-local prefill and decode tokens. The phases are co-batched through TP-sharded linear operators, diverge during attention, and reconverge before the output projection: decode remains localized within its logical rank, while prefill applies Ulysses-SP across corresponding TP ranks of different logical ranks. Row-parallel output and FFN projections produce partial outputs that are reconstructed by the base-TP \texttt{All-Reduce} collectives.
}
    \label{fig:novel-parallelism-tp}
\end{figure*}

The main discussion of \para assumes one GPU per logical rank. For models that require TP to fit model weights or provide sufficient KV-cache capacity, \para instead treats each fixed TP group as one logical rank. Figure~\ref{fig:novel-parallelism-tp} illustrates this composition with base TP2 and two logical ranks, for a total of four GPUs. Prefill applies SP across the two logical ranks, while each decode request remains localized within one logical rank. Both phases retain the same base-TP layout within each rank.

\paragraph{Shared linear operators.}
At the beginning of each Transformer block, the rank-local prefill and decode tokens are replicated across the GPUs within their base-TP group. The QKV projection follows standard column-parallel TP: each TP rank applies its local QKV weight shard and produces the corresponding subset of Q/K/V heads. Prefill and decode tokens can therefore remain co-batched through the projection because they share the same rank-local input and base-TP weight partition.

\paragraph{Phase-specific attention.}
The two phases diverge after the QKV projection. Decode attention remains entirely within its logical rank: each TP rank computes attention for its local decode head shard using the KV cache resident in that logical rank. Prefill additionally applies Ulysses-SP across logical ranks. For each base-TP shard, the corresponding TP ranks from different logical ranks form an independent SP group. The first \texttt{All-to-All} transforms the prefill activations from sequence partitioning across logical ranks into finer head partitions, allowing each physical GPU to compute its assigned prefill attention heads for all current-chunk queries using the cached history and current-chunk K/V. A second \texttt{All-to-All} restores the sequence-partitioned layout while preserving the base-TP partition.

\paragraph{Rejoining the shared path.}
After attention, the prefill and decode paths reconverge on each TP rank and are co-batched through the TP-sharded output projection. Because the output projection is row-parallel, each TP rank produces a full-shape partial output, which is summed across the base-TP group with an \texttt{All-Reduce}. The resulting hidden states are again replicated within the TP group. The FFN follows the standard TP pattern: its expansion projections are column-parallel, the contraction projection is row-parallel, and a second \texttt{All-Reduce} reconstructs the full hidden state.

Thus, composing \para with TP does not require the two phases to share the same parallelism across logical ranks. The base-TP configuration is common to both phases and its collectives remain unchanged, while \para applies additional SP only to the prefill attention path across logical ranks. This preserves prefill--decode co-batching in the shared linear operators while allowing distributed prefill attention and localized decode attention to coexist over the same base-TP execution layout.

\section{Latency Model Details}
\label{app:latency-model}

This section details the base latency functions used by the estimator in \S\ref{sec:estimator}. Following prior operator-level latency models~\cite{agrawal2024vidur,zhong2024distserve,wu2024loongserve}, we model each component as a function of its dominant workload metric and fit the corresponding coefficients from offline execution profiles.

\paragraph{Shared linear operators.}
The shared QKV/O projections and FFNs process the rank-local prefill and decode tokens together. We model their per-layer latency as
\begin{equation}
f_{\scriptscriptstyle \mathrm{GEMM}}(n)
=
\gamma n+\gamma_0,
\end{equation}
where $n$ is the number of rank-local tokens. The fitted coefficients absorb fixed execution overheads, including communication internal to the base TP configuration when applicable.

\paragraph{Decode attention.}
For localized decode attention, the dominant workload is proportional to the aggregate context length of the resident requests. We model
\begin{equation}
f_D(W)=a_D W+a_{D,0},
\end{equation}
where $W=\sum_i c_i$ is the aggregate decode-attention workload on a rank.

\paragraph{Prefill attention.}
For a prefill chunk of length $\ell_i$ with existing context $c_i$, the causal-attention workload is proportional to
\begin{equation}
\ell_i\left(c_i+\frac{\ell_i+1}{2}\right).
\end{equation}
We therefore model the prefill-attention latency as
\begin{equation}
f_P(W)=a_P W+a_{P,0},
\end{equation}
where $W$ denotes the per-rank prefill-attention workload after Ulysses-SP redistributes the scheduled prefill work across ranks.

\paragraph{SP communication.}
The two Ulysses \texttt{All-to-All} collectives communicate the QKV states before attention and the output states afterward. Their latency is modeled from the largest initial prefill shard:
\begin{equation}
f_{\mathrm{comm}}(n)
=
a_{\mathrm{comm}} n+a_{\mathrm{comm},0},
\end{equation}
where $n=\max_r n_p^r$. Equivalently, the bandwidth-dominated component scales with
\begin{equation}
n\frac{B_{\mathrm{QKV}}+B_O}{\mathit{BW}},
\end{equation}
where $B_{\mathrm{QKV}}$ and $B_O$ denote the per-token communication volumes and $\mathit{BW}$ the effective interconnect bandwidth.

\paragraph{Fitting.}
We obtain the base functions by profiling shared-linear execution, prefill attention, decode attention, and SP communication independently over their relevant workload ranges. We then profile heterogeneous \para batches to calibrate the contention factors $\alpha$ and $\beta$ used in Eq.~(\ref{eq:attn_boundary}). This separates the standalone component costs from the slowdown introduced by concurrent execution.

\section{Waterfill Dispatch}
\label{app:dispatch}

\begin{algorithm}[tb]
\caption{Waterfill Dispatch}
\label{alg:waterfill}
\small
\begin{algorithmic}[1]

\Require TBT SLO $\tau$, decode state $\mathcal D$, selected prefills $\mathcal P$, existing prefill assignment $\mathbf A_{\mathrm p}$, optional request $i$, alignment quantum $q$
\Ensure Prefill budget $B$ if $i=\emptyset$; otherwise, prefill chunk $\ell_i$ and its per-rank assignment $\mathbf A_i$

\State $\triangleright$ Find the largest feasible water level
\State $n_{\mathrm{target}}\gets\Call{BinarySearch}{\tau,\mathcal D,\mathcal P,\mathbf A_{\mathrm p},i,q}$
\State $B\gets\sum_{r=1}^{N}\max\{0,n_{\mathrm{target}}-n_{\mathrm d}^r-n_{\mathrm p}^r(\mathbf A_{\mathrm p})\}$

\State \textbf{if} $i=\emptyset$ \textbf{then return} $B$

\State $\ell_i\gets\min\{m_i,B\}$
\State $\mathbf A_i\gets\Call{WaterfillAssign}{\mathcal D,\mathbf A_{\mathrm p},\ell_i,n_{\mathrm{target}}}$

\State \textbf{return} $\ell_i,\mathbf A_i$

\end{algorithmic}
\end{algorithm}

Algorithm~\ref{alg:waterfill} details the waterfill dispatcher used by \sys. The dispatcher takes the current decode state $\mathcal D$, the selected prefill chunks $\mathcal P$, their existing per-rank assignment $\mathbf A_{\mathrm p}$, and an optional request $i$. It uses binary search with the latency estimator to find the largest feasible combined-load target $n_{\mathrm{target}}$ under the TBT SLO $\tau$. When request $i$ is provided, this target is request-specific because the latency contribution of the candidate chunk depends on its current context length. To account for kernel quantization effects, \sys restricts $n_{\mathrm{target}}$ to a hardware-dependent alignment quantum $q$ determined empirically.

Given $n_{\mathrm{target}}$, the additional prefill capacity available on rank $r$ is
\[
\max\{0,n_{\mathrm{target}}-n_d^r-n_p^r(\mathbf A_{\mathrm p})\},
\]
yielding an aggregate additional budget
\[
B=
\sum_{r=1}^{N}
\max\{0,n_{\mathrm{target}}-n_d^r-n_p^r(\mathbf A_{\mathrm p})\}.
\]

When request $i$ is provided, the dispatcher selects the largest chunk that fits within this budget,
\[
\ell_i=\min\{m_i,B\},
\]
where $m_i$ is the request's remaining prefill length. \textsc{WaterfillAssign} then distributes these $\ell_i$ tokens across ranks toward $n_{\mathrm{target}}$, starting from the existing assignment $\mathbf A_{\mathrm p}$, and returns the incremental per-rank assignment $\mathbf A_i$. The scheduler commits $\mathbf A_i$ only after any required prefill-to-decode transition is validated.

When $i=\emptyset$, the dispatcher omits the candidate-specific prefill workload and returns $B$ as a reference residual budget under the current decode and prefill state. The SLO-aware scheduler uses this mode to obtain $B_0$ for Moore--Hodgson admission.

\section{Dual-Mode Decode Placement}
\label{app:placement}

\begin{algorithm}[tb]
\caption{Dual-Mode Decode Placement}
\label{alg:threshold-placement}
\small
\begin{algorithmic}[1]

\Require Request $i$ completing prefill, decode state $\mathcal D$, KV-utilization threshold $\eta$, KV headroom $h$
\Ensure Selected rank $g^\star$ or \textsc{Infeasible}

\State $\mathcal G\gets$ ranks feasible for $i$ under $\mathcal D$ with KV headroom $h$
\State \textbf{if} $\mathcal G=\emptyset$ \textbf{then}
\IState \textbf{return} \textsc{Infeasible}

\State $\mathcal G_{\mathrm{low}}\gets$ ranks in $\mathcal G$ with KV utilization below $\eta$
\State \textbf{if} $\mathcal G_{\mathrm{low}}\neq\emptyset$ \textbf{then}
\IState \textbf{return} \Call{WorstFit}{$\mathcal G_{\mathrm{low}}$}

\State \textbf{return} \Call{BestFit}{$\mathcal G$}

\end{algorithmic}
\end{algorithm}

Algorithm~\ref{alg:threshold-placement} details the dual-mode decode-placement policy used by \sys. When request $i$ completes prefill, \sys first identifies the set of ranks $\mathcal G$ that can accommodate its KV state while preserving a headroom of $h$. The scheduler passes its tentative decode state $\widetilde{\mathcal D}$, including placements reserved earlier in the same iteration. If no such rank exists, the transition is infeasible and the scheduler does not admit the completing prefill chunk.

When at least one feasible rank has KV utilization below the threshold $\eta$, \sys uses \textsc{WorstFit} over these low-utilization ranks. \textsc{WorstFit} selects the rank with the smallest aggregate decode context, which also corresponds to the lowest KV utilization. This spreads decode workload across ranks, reducing decode-attention imbalance and preserving a high TBT-safe prefill-dispatch budget. Once all feasible ranks reach the threshold, \sys switches to \textsc{BestFit}, selecting the feasible rank with the highest KV utilization that can still accommodate the request. This packs KV state more tightly and preserves free capacity on other ranks for subsequent placements.

\section{Moore--Hodgson Admission}
\label{app:admission}

\begin{algorithm}[tb]
\caption{Moore--Hodgson Admission}
\label{alg:mh-admission}
\small
\begin{algorithmic}[1]
\Require Pending prefills $\mathcal R$, per-iteration token budget $B_0$, time $t$, TBT SLO $\tau$
\Ensure Admitted requests $\mathcal R_{\mathrm{adm}}$, rejected requests $\mathcal R_{\mathrm{rej}}$

\State Sort $\mathcal R$ in EDF order
\State $C\gets0$;
$\mathcal R_{\mathrm{rej}}\gets\emptyset$
\State $\mathcal H\gets\textsc{MaxHeap}(\text{key}=m_i)$

\State \textbf{for each} $i\in\mathcal R$
\IState $C\gets C+m_i$;
$\mathcal H.\textsc{push}(i)$
\IState $K \gets
\max\left\{
0,\;
\left\lfloor\frac{d_i-t}{\tau}\right\rfloor B_0
\right\}$
\Comment{Capacity before $d_i$}
\IState \textbf{if} $C>K$ \textbf{then}
\IIState $j\gets\mathcal H.\textsc{popMax}()$;
$C\gets C-m_j$
\IIState $\mathcal R_{\mathrm{rej}}\gets\mathcal R_{\mathrm{rej}}\cup\{j\}$

\State $\mathcal R_{\mathrm{adm}}\gets
\mathcal R\setminus\mathcal R_{\mathrm{rej}}$
\State \textbf{return} $\mathcal R_{\mathrm{adm}},\mathcal R_{\mathrm{rej}}$
\end{algorithmic}
\end{algorithm}

Algorithm~\ref{alg:mh-admission} details the TTFT-aware admission procedure used by \sys. Let $m_i$ denote the remaining prefill length of request $i$, and let $d_i$ denote its TTFT deadline. The dispatcher first provides a reference per-iteration prefill budget $B_0$ under the current decode state. Under this fixed-capacity surrogate, the cumulative prefill capacity available before deadline $d_i$ is
\[
K_i=\max\left\{0,\left\lfloor\frac{d_i-t}{\tau}\right\rfloor B_0\right\},
\]
where $t$ is the current time and $\tau$ is the TBT SLO.

Requests are processed in EDF order while maintaining the total remaining prefill work $C$ of the currently admitted requests. Whenever adding request $i$ causes $C>K_i$, the Moore--Hodgson algorithm~\cite{moore1968} removes the admitted request with the largest remaining prefill length using a max-heap. Since $K_i$ is nondecreasing in EDF order, this single removal restores feasibility before proceeding to the next request.

The surviving requests form $\mathcal R_{\mathrm{adm}}$ and retain EDF priority in the scheduler. Removed requests form $\mathcal R_{\mathrm{rej}}$ and are subsequently considered in SJF order for opportunistic backfill.

\end{document}